\documentclass[prl,amsmath,amssymb,superscriptaddress,twocolumn]{revtex4-1}
\usepackage{graphicx}
\usepackage{dcolumn}
\usepackage{epsfig}
\usepackage{float}
\usepackage{hyperref}
\usepackage[bbgreekl]{mathbbol}

\usepackage{mathrsfs}

\usepackage{lipsum}
\usepackage[cal=boondox,scr=boondoxo]{mathalfa}

\usepackage{trsym} 

\hypersetup{
    colorlinks=true,
    linkcolor=blue,
    filecolor=magenta,      
    urlcolor=cyan,
    pdftitle={An Example},
    pdfpagemode=FullScreen,
    }

\def\Tr{\mathrm{Tr}}

\def\id{\mathbb{1}} 

\usepackage[T3,T1]{fontenc}
\DeclareSymbolFont{tipa}{T3}{cmr}{m}{n}
\DeclareMathAccent{\invbreve}{\mathalpha}{tipa}{16}

\usepackage{accents}
\newlength{\hhatheight}

\usepackage{xcolor}
\usepackage{soul}
\usepackage{changes}

\definechangesauthor[name=Yantao, color=green]{y}
\definechangesauthor[name=Ben, color=brown]{b}
\definechangesauthor[name=Max, color=orange]{m}

\makeatother

\begin{document}

\title{Topological Mixed Valence Model in Magic-Angle Twisted Bilayer Graphene}

\author{Yantao Li}
\affiliation{Department of Physics, Kent State University, Kent, Ohio 44242, USA}

\author{Benjamin M. Fregoso}
\affiliation{Department of Physics, Kent State University, Kent, Ohio 44242, USA}

\author{Maxim Dzero}
\affiliation{Department of Physics, Kent State University, Kent, Ohio 44242, USA}

\begin{abstract}
We develop a model to describe the mixed valence regime in magic-angle twisted bilayer graphene (MATBG) using the recently developed heavy-fermion framework. By employing the large-$N$ slave-boson approach, we derive the self-consistent mean field equations and solve them numerically. We find that the SU(8) symmetry constraint moir\'e system exhibits novel mixed-valence properties which are different from conventional heavy-fermions systems. We find the solutions describing the physics at the filling near the Mott insulator regime in the limit of strong Coulomb interactions between the flat-band fermions. Our model can provide additional insight into the possible microscopic origin of unconventional superconductivity in MATBG.
\end{abstract}

\date{\today}

{
\let\clearpage\relax
\maketitle
}

\emph{Introduction}.---%
The discovery of correlated electronic phases including superconductivity in magic-angle twisted bilayer graphene (MATBG)~\cite{cao2018correlated,cao2018unconventional} has stimulated research efforts to explore various electronic properties in graphene-based multilayer structures~\cite{yankowitz2019tuning,Sharpe_2019,Fleischmann_2020,Jones_2020,cao2021large,Xu_2021,Choi_2021,Park2022,uri2023superconductivity} as well as in van der Waals heterostructures and other platforms~\cite{Naik_2018,Tang2020,Wang2020,Zhang_2020,Shabani_2021,Xu2022,Xiong2022,Meng2023}. As a result, a new field "twistronic physics" has emerged~\cite{Kennes2021} which focuses on theoretical aspects of these systems and covers both static~\cite{Xu_2018,Po_2018,Fengcheng_2018,Kang_2018,Kang_2019,Koshino_2019,Hejazi_2019,Chebrolu_2019,Cea_2019,Liu_2019,Liang_2020,Tran_2020,Tritsaris_2020,Ramires2021,Lake2021,Qin2021,Eaton2022,Zhou2022} and non-equilibrium properties~\cite{Topp_2019,Li_2020,Katz_2020,Vogl2020,Vogl2021}. 

The MATBG system consists of two single graphene sheets which are twisted relative to each other at certain angles called magic angles~\cite{Bistritzer_2011,dosSantos_2007,Shallcross_2008,Shallcross_2010,Mele_2010,Mele_2011,Carr_2017}. It is believed that it is the flat bands that appear at such magic angles are the main driver for the exotic physical phenomena which were experimentally observed in these systems. The origin of magic angles is related to the case when the electron tunneling in the AA-region of MATBG is neglected corresponding to the chiral limit~\cite{Tarnopolsky_2019,Khalaf_2019}. It can be theoretically shown that one does not need to use the chiral limit for other twisted graphene stacks to exhibit such flat bands associated with Dirac cones ~\cite{Li2022}. Furthermore, the discovery of superconductivity in MATBG demonstrates yet another example of superconductivity emerging from the 'strange metal' phase~\cite{Cao2020S} and, as such, is reminiscent of the physics of the high-Tc copper-based and heavy-fermion superconductors. In passing we note, that several works have recently attempted to explain the origin of superconductivity in MATBG from various perspectives~\cite{Wu2018super,Lian2019,Gonzalez2019,Christos2022,Fischer2022}.

Most recently an alternative viewpoint has emerged. Specifically, focusing on a first magic angle, Song and Bernevig showed that a model for twisted bilayer graphene can be mapped to the heavy-fermion model ~\cite{Song2022}. This elegant theory is based on the experimental fact that AA-region in MATBG exhibits a quantum dot-like behavior~\cite{Tilak2021} and so one can describe the physics of this region using a model with the flat band electrons ($f$-electrons). The electrons in the AB/BA regions play the role of conduction electrons ($c$-electrons). By mapping the Bistritzer-MacDonald model~\cite{Bistritzer_2011} to the periodic Anderson model at the first magic angle, Song and Bernevig created a way to bridge the MATBG with the more conventional heavy-fermion systems. However, in striking contrast with the conventional heavy-fermion system, the MATBG hosts SU(8) symmetry due to spin and valley degrees of freedom and two central flat bands. Shortly after several important works appeared, which addressed various aspects of this unique moir\'e  heavy-fermion-like system~\cite{chou2022kondo,yu2023magicangle,hu2023symmetric,hu2023kondo,zhou2023kondo,huang2023evolution,calugaru2023tbg,singh2023topological,chou2023scaling}. 

In conventional heavy-fermion systems, one usually distinguishes between the so-called local moment (or Kondo) regime and the mixed-valent one. In the local moment regime, the energy of the flat ($f$-orbital) band lies well below the Fermi energy of the conduction electrons, while in the mixed-valent regime, it lies close to the Fermi energy. In order to describe both of these regimes on a technical level, one considers the limit when the local Coulomb repulsion is taken to infinity. Then one introduces the projection operators along with the Lagrange multiplier to enforce the constraint of the single occupancy on the $f$-levels. In the mean-field approximation, one replaces the projection operators and constraint fields with the $c$-numbers which are computed self-consistently and describe the renormalization of the $f$-energy level, hybridization between $c$ and $f$-electrons as well as renormalization of the chemical potential. 

Early works on the application of the heavy-fermion model to MATBG have typically focused on the Kondo regime and have not solved the problem self-consistently. In this paper, we attempt to resolve this issue. In what follows, we present a new heavy-fermion model for the MATBG and discuss the physics both in the Kondo regime and the mixed valence regime self-consistently. Our self-consistent solutions pave the way to go beyond the mean-field approximation and consider the effects of fluctuations on the competing phases of MATBG. 

An important aspect of our model is that we consider the strong limit of Coulomb interactions, this could provide another way to explain the microscopic origin of unconventional ($d$-wave) superconductivity, which can be shown to arise from the quantum mechanical fluctuations in the number of the $f$-electrons and is purely electronic in origin. Indeed, as the very recent experiment suggests~\cite{Oh2021}, the MATBG does indicate the $d$-wave superconductivity similar to the Cooper pair symmetry of the superconductivity in the conventional heavy fermion systems~\cite{Lavagna1987}.

\emph{Setup and Topological Heavy Fermion Model}.---%
The MATBG setup is illustrated in Fig.~\ref{Fig1}(a). Two single graphene layers are stacked together and they are twisted relative to each other with angle $\theta$. The AA regions are designated by white spots (with red arrows depicting flat-band electrons) and AB/BA regions are dark green (with blue spheres depicting conduction electrons). 

Next, we consider the first magic angle $\theta_m=1.05^{\circ}$. At this magic angle, the MATBG can be mapped from the Bistritzer-MacDonald model to a special SU(8) periodic Anderson model, i.e. the Song-Bernevig model (SB)~\cite{Song2022}.  We choose the SB model as the starting point. The Hamiltonian is 
\begin{equation}
\hat{H}=\hat{H}_{0,\textrm{c}}+\hat{H}_{0,\textrm{f}}+\hat{H}_{0,\textrm{cf}}+\hat{H}_{\textrm{U}},
\end{equation}
where 
\begin{equation}
	\hat{H}_{0\textrm{c}}=\sum_{a,a^{\prime},\eta,s}\sum_{\bold{p}} h^{(c,\eta s)}_{a a^{\prime}}(\bold{p}) c^{\dagger}_{\bold{p},a,\eta,s} c_{\bold{p},a^{\prime},\eta,s},
\end{equation}
is the Hamiltonian of the conduction ($c$-) electrons in AB/BA moir\'e lattice sites, 
\begin{equation}
	\hat{H}_{0\textrm{f}}=\sum_{\alpha,\alpha^{\prime},\eta,s}\sum_{\bold{k}} h^{(f,\eta s)}_{\alpha \alpha^{\prime}}(\bold{k}) f^{\dagger}_{\bold{k},\alpha,\eta,s} f_{\bold{k},\alpha^{\prime},\eta,s},
\end{equation}
is the Hamiltonian of flat band ($f$-) electrons in AA moir\'e lattice sites, 
\begin{equation}
\begin{aligned}
	\hat{H}_{0\textrm{cf}}=&\sum_{\alpha,a,\eta,s}\sum_{\bold{G}}\sum_{\bold{k}\in \text{mBZ}} [V^{(\eta s)}_{\alpha a}(\bold{k}+\bold{G}) f^{\dagger}_{\bold{k},\alpha,\eta,s} c_{\bold{k}+\bold{G},a,\eta,s}\\&+\text{H.c.}],
\end{aligned}	
\end{equation}
accounts for the hybridization between the $c$ and $f$-electrons
and
\begin{equation}
	\hat{H}_{\textrm{U}}=\frac{U}{2}\sum_{\bold{R}}:\hat{n}^{f}_{\bold{R}}::\hat{n}^{f}_{\bold{R}}:,
\end{equation}
is the Hamiltonian describing the local Coulomb repulsion between the $f$- electrons. 
\begin{figure}[]
   \centering
   \includegraphics[width=3.4in]{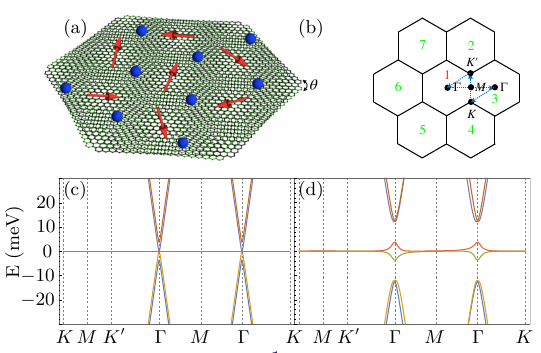} 
   \caption{(a) Sketch of the setup. Two single graphene layers twist relative to each other with angle $\theta=\theta_m$. The AA regions behave like quantum dots (red arrows) and AB/BA regions behave like conduction states (blue balls). (b) Moir\'e momentum space with 7 mBZs. The blue dashed lines mean the path of the spectrum. (c) and (d) are spectrum with parameters $\rho=0$ (before hybridization) and $\rho=0.5$ (full hybridization) separately. Also, both of them are set with $\mu_f=\lambda=0$ and other parameters can be seen in the main text.}
   \label{Fig1}
\end{figure}
In the expressions above $\alpha=1,2$, $a=1,2,3,4$, $\eta=\pm$, and $s=\uparrow,\downarrow$ are flat band, conduction band, valley, and spin indexes correspondingly, $\bold{p}=\bold{k}+\bold{G}$, $\bold{G}$ is the reciprocal lattice vectors in the moir\'e momentum space, $\hat{n}^{f}_{\bold{R}}$ is the on-site density operator of $f$-electrons, $U$ is the strength of the Coulomb repulsion. Lastly, the matrices which appear in the expressions above are defined according to
\begin{equation}
\begin{split}
\hat{h}^{(c,\eta s)}(\bold{p})&=\left[\begin{matrix}
-\mu_c\hat{\sigma}_0 & \nu_\star(\eta p_x\hat{\sigma}_0+ip_y\hat{\sigma}_z) \\
\nu_\star(\eta p_x\hat{\sigma}_0-ip_y\hat{\sigma}_z)  & M\hat{\sigma}_x-\mu_c\hat{\sigma}_0\end{matrix}\right], \\
V^{(\eta s)}(\bold{p})&=e^{\frac{-|\bold{p}|^{2}\lambda^{2}}{2}}
\begin{bmatrix}
\gamma\sigma_0+\nu_\star^{\prime}(\eta p_x\sigma_x+p_y\sigma_y)\\
0_{2\times2}\\
\end{bmatrix},
\end{split}
\end{equation}
and $h^{(f,\eta s)}=(\epsilon_{f_0}-\mu_f)\hat{\sigma}_0$, where $\mu_c$ and $\mu_f$ are the chemical potentials for $c$ and $f-$ electrons accordingly, $\hat{\sigma}_0=I_{2\times 2}$ is the unit matrix, $\hat{\sigma}_j$ ($j=x,y,z$) are the Pauli matrices. Note that the Hamiltonian is expressed in the moir\'e momentum space using the plane wave approximation. The size of moir\'e momentum space (per valley per spin) is $2+4N_G$ which means $2$ flat bands and $4$ conduction bands with $N_G$ moir\'e Brillouin zones (mBZ). The values of the parameters are $\nu_\star=-4.303$ eV\AA, $M=3.697$ meV, $\gamma=-24.75$ meV, $\nu_\star^{\prime}=1.622$ eV\AA, and the dampling factor $\lambda=0.3375$ $a_{M}$, where $a_{M}$ is the moir\'e lattice constant. Note that all these parameters correspond to 'magic angle' $\theta_m=1.05^{\circ}$, $U_0=W_{AA}/W_{AB}=0.8$, $W_{AA}$ and $W_{AB}$ are interlayer hopping amplitudes in AA regions, and AB/BA regions separately, and the velocity of the electron in single layer graphene $v_F=5.94$ eV\AA. We note that when $M=0$ the system reaches the limit of the flat band and we do \textit{not} consider this case in this paper.

\emph{Slave-Boson approach and Mean Field Equations}.---%
To handle the SB model in the mixed valence region, we use the slave-boson approach. We extend the number of orbital, spin, and valley for both $c$- and $f$-electrons to $N$ flavors and set the interaction
\begin{equation}
	U=\infty 
\end{equation}
to exclude the double occupancy. Introducing the slave-boson operators $b^{\dagger}_{\bold{R}}$ and $b_{\bold{R}}$ at each AA site in real space  the constraint becomes 
\begin{equation}
	Q=\sum^{N}_{l=1}\sum_{\alpha}(f^{\dagger}_{\bold{R},\alpha, l} f_{\bold{R},\alpha, l}+b^{\dagger}_{\bold{R}}b_{\bold{R}}).
\end{equation}
Note that $N=4$ includes spin and valley for two flat band orbitals. Here, we suppose different valley has the same band structure and we choose the valley $\eta=+$ in the above Hamiltonian. We stress that there are two index spaces in the Hamiltonian. One is the extended $N$ space with $1/N$ expansion corresponding to the index $l$. Another is the moir\'e momentum space which has a size of $2+4N_G$ corresponding to the index $\alpha$, $\alpha^{\prime}$, $a$, and $a^{\prime}$ separately 

We introduce Lagrangian multipliers $\lambda_{\bold{R}}$ to ensure the number of $f$ electrons is $Q=1$. We rewrite $Q\rightarrow q_0 N$, $b_{\bold{R}}\rightarrow b_{\bold{R}} \sqrt{N}$, and $V_{\alpha a}\rightarrow V_{\alpha a}/ \sqrt{N}$. The local gauge transformation is $b_{\bold{R}}=\rho_{\bold{R}}\exp(i\theta_{\bold{R}})$, $f_{\bold{R}}=f^{\prime}_{\bold{R}}\exp(i\theta_{\bold{R}})$ and $\lambda_{\bold{R}}=\lambda^{\prime}_{\bold{R}}-\theta_{\bold{R}}$. We then rewrite $f^{\prime}_{\bold{R}}$ and $\lambda^{\prime}_{\bold{R}}$ to $f_{\bold{R}}$ and $\lambda_{\bold{R}}$. The partition function is 
\begin{equation}
      Z=\int \mathscr{D}(cc^{\dagger}ff^{\dagger}\rho\lambda)\exp(-S),	
\end{equation}
where the action is $S=\int_0^{\beta} L(\tau) d\tau$, and
\noindent
\begin{widetext}
\begin{equation}
\begin{aligned}
	& L
	=\sum^{N}_{\substack{a,a^{\prime};l=1\\
                  \bold{p}}}(\partial_{\tau}+ h^{(c)}_{a a^{\prime}}(\bold{p})) c^{\dagger}_{\bold{p},a, l}c_{\bold{p},a^{\prime},l}(\tau)
	+\sum_{\substack{\alpha,\alpha^{\prime}\\
                  \bold{\bold{k},\bold{k}^{\prime}\in \text{mBZ}}}}(\partial_{\tau}+h^{(f)}_{\alpha \alpha^{\prime}}(\bold{k})\delta_{\bold{k},\bold{k}^{\prime}}+\frac{i\lambda(\bold{k}-\bold{k}^{\prime};\tau)}{\sqrt{N_{L}}}) f^{\dagger}_{\bold{k},\alpha, l} f_{\bold{k}^{\prime},\alpha^{\prime}, l}(\tau)+\frac{1}{\sqrt{N_{L}}}\sum^{N}_{\substack{\bold{G},\alpha,a;l=1\\
                  \bold{\bold{k},\bold{k}^{\prime}\in \text{mBZ}}}}
	\\&\left.[V_{\alpha a}(\bold{k}+\bold{G}) c^{\dagger}_{\bold{k}+\bold{G},a, l}(\tau)f_{\bold{k}^{\prime},\alpha, l}(\tau)\rho^{\dagger}(\bold{k}-\bold{k}^{\prime};\tau)+\text{H.c.}]\right.
	+\left.\frac{iN}{\sqrt{N_{L}}}\sum_{\bold{k},\bold{k}^{\prime}\in \text{mBZ}}  \rho({\bold{k}};\tau)\lambda(\bold{k}^{\prime}-\bold{k};\tau)\rho(-\bold{k}^{\prime};\tau)\right.\left.-iq_0N\sqrt{N_L}\lambda(0;\tau)\right., 
\end{aligned}
\end{equation}       
\end{widetext}
where $\beta=\frac{1}{T}$, $T$ is the temperature, and $N_L$ is the number of lattice site in moir\'e real space. We set $\lambda(\bold{k;\tau})=\frac{1}{T}\bar{\lambda}\delta_{\bold{k},0}$, $\rho(\bold{k};\tau)=\frac{1}{T}\bar{\rho}\delta_{\bold{k},0}$ to get the mean field action. We rewrite $\frac{\bar{\lambda}}{\sqrt{N_{L}}}\rightarrow \bar{\lambda}$ and $\frac{\bar{\rho}}{\sqrt{N_{L}}}\rightarrow \bar{\rho}$. After applying $\partial S_{0}/\partial \bar{\rho}=0$, $\partial S_{0}/\partial \bar{\lambda}=0$, and summing over the Matsubara frequency $\omega_n$, we end up with three mean field equations as follows
\begin{equation}
    \bar{\rho}\bar{\lambda}=\frac{i}{2 N_{L}}\sum_{\bold{k}\in \text{mBZ}}\sum^{2+4N_G}_{j}(P^{\dagger}A^{\bar{\rho}}_0P)_{jj}\cdot n_{F}(\mathcal{E}_j),\label{eq1}
\end{equation}
\begin{equation}
      q_0-\bar{\rho}^{2}=-\frac{i}{N_{L}}\sum_{\bold{k}\in \text{mBZ}}\sum^{2+4N_G}_{j}(P^{\dagger}A^{\bar{\lambda}}_0P)_{jj}\cdot n_F(\mathcal{E}_j),\label{eq2}
\end{equation}
\begin{equation}
      n_t=q_0-\bar{\rho}^{2}+\frac{1}{N_{L}}\sum_{\bold{k}\in \text{mBZ}}\sum^{2+4N_G}_{j}(P^{\dagger}A^{c}_0P)_{jj}\cdot n_F(\mathcal{E}_j)-2N_G,\label{eq3}
\end{equation}
where $n_t$ is the total filling, $n_F(\epsilon)=(\exp(\epsilon/T)+1)^{-1}$ is the Fermi-Dirac distribution, $\mathcal{E}_j$ is the eigenvalues of matrix $A_0=\{\{\hat{h}^{(c)},\hat{V}\},\{\hat{V}^{\dagger}, \hat{h}^{(f)}+i\bar{\lambda}\sigma_0 \}\}$, $A^{\bar{\rho}}_0=\partial{(-i\omega_n+A_0)}/\partial \bar{\rho}$, $A^{\bar{\lambda}}_0=\partial{(-i\omega_n+A_0)}/\partial \bar{\lambda}$, $A^{c}_0=-\partial{(-i\omega_n+A_0)}/\partial \mu_c$, and $A_0$ expands in the moir\'e momentum space. Note that $P$ can be constructed by the eigenvectors of $A_0$ and $P=(c_1,c_2,\cdots,c_{(2+4N_G)})_{2+4N_G\times2+4N_G}$, where $c_j$ are the eigenvectors of $A_0$. (See Supplemental Material~\cite{Note1} for the detailed derivation). The above three self-consistent mean field equations are one of our main results.

\emph{Numerics}.---%
Now, we will numerically solve the equations Eq.~\ref{eq1}, Eq.~\ref{eq2}, and Eq.~\ref{eq3}. To solve them self-consistently, we set an error bar $errs=\sum^{3}_{n=1}(l_n-r_n)^{2}$, where $l_n$ and $r_n$ represent left and right sides of $n$th mean field equation separately. We set $\mu_c=\mu_f$ and go through the parameters regions $\rho \in[0,0.5]$, $\mu_f \in [-100,100]$ meV, and $i\lambda \in [-100,100]$ meV. Since we set the interaction $U=\infty$, $Q=1$, so we have $q_0=1/4$. To reach the mixed valence region, we also set the total filling $n_t=0.83\times q_0$. We find the solutions to make the $errs\approx0$. 
There exist two solutions: one is positive $\mu_f$, and another is negative $\mu_f$. We note that $\mathcal{E}_j(\bar{\lambda},\bar{\rho},\mu_f,\bold{k})$ is numerically calculated and depends on the chemical potential $\mu_f$ and momentum $\bold{k}$ with $\bold{k}\in \text{mBZ}$.

We substitute the solutions to $A_0$ and we get the spectrums, see Fig.~\ref{Fig2}. We also plot the variation of parameters in the mixed valence region as a function of the temperature $T$, see Fig.~\ref{Fig2}.
\begin{figure}[]
   \centering
   \includegraphics[width=3.4in]{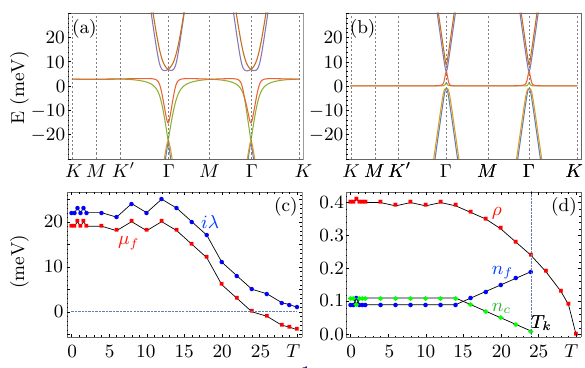} 
   \caption{(a) Spectrum of the self-consistent solutions with parameters $\mu_f=19$ meV, $i\lambda=22$ meV, $\rho=0.4$, and $T=0.01K$. (b) The spectrum of the self-consistent solutions with parameters $\mu_f=-5$ meV, $i\lambda=-5$ meV, $\rho=0.1$, and $T=0.01K$. (c) Self-consistent solutions $\mu_f$ and $i\lambda$ vs $T$. (d) Self-consistent solutions $\rho$, $n_f$, and $n_c$ vs $T$ with $n_t=0.8q_0$, where $T_k\approx24K$ is the Kondo temperature.}
   \label{Fig2}
\end{figure}

\emph{Discussion}.---%
Different from the conventional heavy fermion system, the topological mixed valence model in MATBG has SU(8) symmetry. The two central flat bands for each valley and spin are particle-hole symmetric in the chiral limit. Although the particle-hole symmetry is broken at the experimental lattice relaxation range $U_0=0.8$, the valence bands host a similar band structure as the conduction bands before the hybridization. This makes the difficult to reach the Kondo region just by pushing the flat bands lower away from the conduction bands as it does in the conventional heavy fermion system. One needs to consider many valence bands together. We also note that it might be interesting to consider different flavors of slave boson~\cite{Dorin1993} since we have a total of 8 flat bands; they are not degenerate at the experimental range. We leave this task for future work.

\emph{Summary}.---%
We have introduced a new model to describe the mixed valence region of the magic-angle twisted bilayer graphene with infinite Coulomb interaction. We start from the SB model and use the slave-boson method in large-$N$ expansion. We derive a new group of mean field equations to describe the mixed valence regions of twisted bilayer graphene. The solutions can catch the physics of the filling near the strong correlation, which is at the edge of the Mott insulator, and then could be approaching the unconventional superconductivity. Our topological mixed valence model paves the way to study the possible origin of superconductivity in twisted bilayer graphene. We hope our model could stimulate further research in the mixed valence region in various related van der Waals heterostructures materials or platforms.

\emph{Note added}.---%
A related paper comes out recently~\cite{lau2023topological}, which deals with the mixed valence model in twisted bilayer graphene with finite Coulomb interaction.
\begin{acknowledgments}
\emph{Acknowledgments.} We would like to acknowledge the very useful discussions with Yang-Zhi Chou. This work was financially supported by the National Science Foundation Grants No. NSF-DMR-2002795 (Y. L. and M. D.) and NSF-DMR-2015639 (B.M.F.). Parts of this paper were written during the Aspen Center of Physics 2023 summer program on ``New Directions on Strange Metals in Correlated Systems" (M. D.), which was supported by the National Science Foundation Grant No. PHY-2210452.
\end{acknowledgments}

\bibliography{ref}

\begin{thebibliography}{74}%
\makeatletter
\providecommand \@ifxundefined [1]{%
 \@ifx{#1\undefined}
}%
\providecommand \@ifnum [1]{%
 \ifnum #1\expandafter \@firstoftwo
 \else \expandafter \@secondoftwo
 \fi
}%
\providecommand \@ifx [1]{%
 \ifx #1\expandafter \@firstoftwo
 \else \expandafter \@secondoftwo
 \fi
}%
\providecommand \natexlab [1]{#1}%
\providecommand \enquote  [1]{``#1''}%
\providecommand \bibnamefont  [1]{#1}%
\providecommand \bibfnamefont [1]{#1}%
\providecommand \citenamefont [1]{#1}%
\providecommand \href@noop [0]{\@secondoftwo}%
\providecommand \href [0]{\begingroup \@sanitize@url \@href}%
\providecommand \@href[1]{\@@startlink{#1}\@@href}%
\providecommand \@@href[1]{\endgroup#1\@@endlink}%
\providecommand \@sanitize@url [0]{\catcode `\\12\catcode `\$12\catcode
  `\&12\catcode `\#12\catcode `\^12\catcode `\_12\catcode `\%12\relax}%
\providecommand \@@startlink[1]{}%
\providecommand \@@endlink[0]{}%
\providecommand \url  [0]{\begingroup\@sanitize@url \@url }%
\providecommand \@url [1]{\endgroup\@href {#1}{\urlprefix }}%
\providecommand \urlprefix  [0]{URL }%
\providecommand \Eprint [0]{\href }%
\providecommand \doibase [0]{http://dx.doi.org/}%
\providecommand \selectlanguage [0]{\@gobble}%
\providecommand \bibinfo  [0]{\@secondoftwo}%
\providecommand \bibfield  [0]{\@secondoftwo}%
\providecommand \translation [1]{[#1]}%
\providecommand \BibitemOpen [0]{}%
\providecommand \bibitemStop [0]{}%
\providecommand \bibitemNoStop [0]{.\EOS\space}%
\providecommand \EOS [0]{\spacefactor3000\relax}%
\providecommand \BibitemShut  [1]{\csname bibitem#1\endcsname}%
\let\auto@bib@innerbib\@empty
\bibitem [{\citenamefont {Cao}\ \emph {et~al.}(2018{\natexlab{a}})\citenamefont
  {Cao}, \citenamefont {Fatemi}, \citenamefont {Demir}, \citenamefont {Fang},
  \citenamefont {Tomarken}, \citenamefont {Luo}, \citenamefont
  {Sanchez-Yamagishi}, \citenamefont {Watanabe}, \citenamefont {Taniguchi},
  \citenamefont {Kaxiras}, \citenamefont {Ashoori},\ and\ \citenamefont
  {Jarillo-Herrero}}]{cao2018correlated}%
  \BibitemOpen
  \bibfield  {author} {\bibinfo {author} {\bibfnamefont {Y.}~\bibnamefont
  {Cao}}, \bibinfo {author} {\bibfnamefont {V.}~\bibnamefont {Fatemi}},
  \bibinfo {author} {\bibfnamefont {A.}~\bibnamefont {Demir}}, \bibinfo
  {author} {\bibfnamefont {S.}~\bibnamefont {Fang}}, \bibinfo {author}
  {\bibfnamefont {S.~L.}\ \bibnamefont {Tomarken}}, \bibinfo {author}
  {\bibfnamefont {J.~Y.}\ \bibnamefont {Luo}}, \bibinfo {author} {\bibfnamefont
  {J.~D.}\ \bibnamefont {Sanchez-Yamagishi}}, \bibinfo {author} {\bibfnamefont
  {K.}~\bibnamefont {Watanabe}}, \bibinfo {author} {\bibfnamefont
  {T.}~\bibnamefont {Taniguchi}}, \bibinfo {author} {\bibfnamefont
  {E.}~\bibnamefont {Kaxiras}}, \bibinfo {author} {\bibfnamefont {R.~C.}\
  \bibnamefont {Ashoori}}, \ and\ \bibinfo {author} {\bibfnamefont
  {P.}~\bibnamefont {Jarillo-Herrero}},\ }\href@noop {} {\bibfield  {journal}
  {\bibinfo  {journal} {Nature}\ }\textbf {\bibinfo {volume} {556}},\ \bibinfo
  {pages} {80} (\bibinfo {year} {2018}{\natexlab{a}})}\BibitemShut {NoStop}%
\bibitem [{\citenamefont {Cao}\ \emph {et~al.}(2018{\natexlab{b}})\citenamefont
  {Cao}, \citenamefont {Fatemi}, \citenamefont {Fang}, \citenamefont
  {Watanabe}, \citenamefont {Taniguchi}, \citenamefont {Kaxiras},\ and\
  \citenamefont {Jarillo-Herrero}}]{cao2018unconventional}%
  \BibitemOpen
  \bibfield  {author} {\bibinfo {author} {\bibfnamefont {Y.}~\bibnamefont
  {Cao}}, \bibinfo {author} {\bibfnamefont {V.}~\bibnamefont {Fatemi}},
  \bibinfo {author} {\bibfnamefont {S.}~\bibnamefont {Fang}}, \bibinfo {author}
  {\bibfnamefont {K.}~\bibnamefont {Watanabe}}, \bibinfo {author}
  {\bibfnamefont {T.}~\bibnamefont {Taniguchi}}, \bibinfo {author}
  {\bibfnamefont {E.}~\bibnamefont {Kaxiras}}, \ and\ \bibinfo {author}
  {\bibfnamefont {P.}~\bibnamefont {Jarillo-Herrero}},\ }\href@noop {}
  {\bibfield  {journal} {\bibinfo  {journal} {Nature}\ }\textbf {\bibinfo
  {volume} {556}},\ \bibinfo {pages} {43} (\bibinfo {year}
  {2018}{\natexlab{b}})}\BibitemShut {NoStop}%
\bibitem [{\citenamefont {Yankowitz}\ \emph {et~al.}(2019)\citenamefont
  {Yankowitz}, \citenamefont {Chen}, \citenamefont {Polshyn}, \citenamefont
  {Zhang}, \citenamefont {Watanabe}, \citenamefont {Taniguchi}, \citenamefont
  {Graf}, \citenamefont {Young},\ and\ \citenamefont
  {Dean}}]{yankowitz2019tuning}%
  \BibitemOpen
  \bibfield  {author} {\bibinfo {author} {\bibfnamefont {M.}~\bibnamefont
  {Yankowitz}}, \bibinfo {author} {\bibfnamefont {S.}~\bibnamefont {Chen}},
  \bibinfo {author} {\bibfnamefont {H.}~\bibnamefont {Polshyn}}, \bibinfo
  {author} {\bibfnamefont {Y.}~\bibnamefont {Zhang}}, \bibinfo {author}
  {\bibfnamefont {K.}~\bibnamefont {Watanabe}}, \bibinfo {author}
  {\bibfnamefont {T.}~\bibnamefont {Taniguchi}}, \bibinfo {author}
  {\bibfnamefont {D.}~\bibnamefont {Graf}}, \bibinfo {author} {\bibfnamefont
  {A.~F.}\ \bibnamefont {Young}}, \ and\ \bibinfo {author} {\bibfnamefont
  {C.~R.}\ \bibnamefont {Dean}},\ }\href@noop {} {\bibfield  {journal}
  {\bibinfo  {journal} {Science}\ }\textbf {\bibinfo {volume} {363}},\ \bibinfo
  {pages} {1059} (\bibinfo {year} {2019})}\BibitemShut {NoStop}%
\bibitem [{\citenamefont {Sharpe}\ \emph {et~al.}(2019)\citenamefont {Sharpe},
  \citenamefont {Fox}, \citenamefont {Barnard}, \citenamefont {Finney},
  \citenamefont {Watanabe}, \citenamefont {Taniguchi}, \citenamefont
  {Kastner},\ and\ \citenamefont {Goldhaber-Gordon}}]{Sharpe_2019}%
  \BibitemOpen
  \bibfield  {author} {\bibinfo {author} {\bibfnamefont {A.~L.}\ \bibnamefont
  {Sharpe}}, \bibinfo {author} {\bibfnamefont {E.~J.}\ \bibnamefont {Fox}},
  \bibinfo {author} {\bibfnamefont {A.~W.}\ \bibnamefont {Barnard}}, \bibinfo
  {author} {\bibfnamefont {J.}~\bibnamefont {Finney}}, \bibinfo {author}
  {\bibfnamefont {K.}~\bibnamefont {Watanabe}}, \bibinfo {author}
  {\bibfnamefont {T.}~\bibnamefont {Taniguchi}}, \bibinfo {author}
  {\bibfnamefont {M.~A.}\ \bibnamefont {Kastner}}, \ and\ \bibinfo {author}
  {\bibfnamefont {D.}~\bibnamefont {Goldhaber-Gordon}},\ }\href {\doibase
  10.1126/science.aaw3780} {\bibfield  {journal} {\bibinfo  {journal}
  {Science}\ }\textbf {\bibinfo {volume} {365}},\ \bibinfo {pages} {605}
  (\bibinfo {year} {2019})}\BibitemShut {NoStop}%
\bibitem [{\citenamefont {Fleischmann}\ \emph {et~al.}(2020)\citenamefont
  {Fleischmann}, \citenamefont {Gupta}, \citenamefont {Wullschl{\"a}ger},
  \citenamefont {Theil}, \citenamefont {Weckbecker}, \citenamefont {Meded},
  \citenamefont {Sharma}, \citenamefont {Meyer},\ and\ \citenamefont
  {Shallcross}}]{Fleischmann_2020}%
  \BibitemOpen
  \bibfield  {author} {\bibinfo {author} {\bibfnamefont {M.}~\bibnamefont
  {Fleischmann}}, \bibinfo {author} {\bibfnamefont {R.}~\bibnamefont {Gupta}},
  \bibinfo {author} {\bibfnamefont {F.}~\bibnamefont {Wullschl{\"a}ger}},
  \bibinfo {author} {\bibfnamefont {S.}~\bibnamefont {Theil}}, \bibinfo
  {author} {\bibfnamefont {D.}~\bibnamefont {Weckbecker}}, \bibinfo {author}
  {\bibfnamefont {V.}~\bibnamefont {Meded}}, \bibinfo {author} {\bibfnamefont
  {S.}~\bibnamefont {Sharma}}, \bibinfo {author} {\bibfnamefont
  {B.}~\bibnamefont {Meyer}}, \ and\ \bibinfo {author} {\bibfnamefont
  {S.}~\bibnamefont {Shallcross}},\ }\href {\doibase
  10.1021/acs.nanolett.9b04027} {\bibfield  {journal} {\bibinfo  {journal}
  {Nano Letters}\ }\textbf {\bibinfo {volume} {20}},\ \bibinfo {pages} {971}
  (\bibinfo {year} {2020})}\BibitemShut {NoStop}%
\bibitem [{\citenamefont {Jones}\ \emph {et~al.}(2020)\citenamefont {Jones},
  \citenamefont {Muzzio}, \citenamefont {Majchrzak}, \citenamefont {Pakdel},
  \citenamefont {Curcio}, \citenamefont {Volckaert}, \citenamefont {Biswas},
  \citenamefont {Gobbo}, \citenamefont {Singh}, \citenamefont {Robinson},
  \citenamefont {Watanabe}, \citenamefont {Taniguchi}, \citenamefont {Kim},
  \citenamefont {Cacho}, \citenamefont {Lanata}, \citenamefont {Miwa},
  \citenamefont {Hofmann}, \citenamefont {Katoch},\ and\ \citenamefont
  {Ulstrup}}]{Jones_2020}%
  \BibitemOpen
  \bibfield  {author} {\bibinfo {author} {\bibfnamefont {A.~J.~H.}\
  \bibnamefont {Jones}}, \bibinfo {author} {\bibfnamefont {R.}~\bibnamefont
  {Muzzio}}, \bibinfo {author} {\bibfnamefont {P.}~\bibnamefont {Majchrzak}},
  \bibinfo {author} {\bibfnamefont {S.}~\bibnamefont {Pakdel}}, \bibinfo
  {author} {\bibfnamefont {D.}~\bibnamefont {Curcio}}, \bibinfo {author}
  {\bibfnamefont {K.}~\bibnamefont {Volckaert}}, \bibinfo {author}
  {\bibfnamefont {D.}~\bibnamefont {Biswas}}, \bibinfo {author} {\bibfnamefont
  {J.}~\bibnamefont {Gobbo}}, \bibinfo {author} {\bibfnamefont
  {S.}~\bibnamefont {Singh}}, \bibinfo {author} {\bibfnamefont {J.~T.}\
  \bibnamefont {Robinson}}, \bibinfo {author} {\bibfnamefont {K.}~\bibnamefont
  {Watanabe}}, \bibinfo {author} {\bibfnamefont {T.}~\bibnamefont {Taniguchi}},
  \bibinfo {author} {\bibfnamefont {T.~K.}\ \bibnamefont {Kim}}, \bibinfo
  {author} {\bibfnamefont {C.}~\bibnamefont {Cacho}}, \bibinfo {author}
  {\bibfnamefont {N.}~\bibnamefont {Lanata}}, \bibinfo {author} {\bibfnamefont
  {J.~A.}\ \bibnamefont {Miwa}}, \bibinfo {author} {\bibfnamefont
  {P.}~\bibnamefont {Hofmann}}, \bibinfo {author} {\bibfnamefont
  {J.}~\bibnamefont {Katoch}}, \ and\ \bibinfo {author} {\bibfnamefont
  {S.}~\bibnamefont {Ulstrup}},\ }\href {\doibase
  https://doi.org/10.1002/adma.202001656} {\bibfield  {journal} {\bibinfo
  {journal} {Advanced Materials}\ }\textbf {\bibinfo {volume} {32}},\ \bibinfo
  {pages} {2001656} (\bibinfo {year} {2020})}\BibitemShut {NoStop}%
\bibitem [{\citenamefont {Cao}\ \emph {et~al.}()\citenamefont {Cao},
  \citenamefont {Park}, \citenamefont {Watanabe}, \citenamefont {Taniguchi},\
  and\ \citenamefont {Jarillo-Herrero}}]{cao2021large}%
  \BibitemOpen
  \bibfield  {author} {\bibinfo {author} {\bibfnamefont {Y.}~\bibnamefont
  {Cao}}, \bibinfo {author} {\bibfnamefont {J.~M.}\ \bibnamefont {Park}},
  \bibinfo {author} {\bibfnamefont {K.}~\bibnamefont {Watanabe}}, \bibinfo
  {author} {\bibfnamefont {T.}~\bibnamefont {Taniguchi}}, \ and\ \bibinfo
  {author} {\bibfnamefont {P.}~\bibnamefont {Jarillo-Herrero}},\ }\href@noop {}
  {\enquote {\bibinfo {title} {Large pauli limit violation and reentrant
  superconductivity in magic-angle twisted trilayer graphene},}\ }\Eprint
  {http://arxiv.org/abs/2103.12083} {arXiv:2103.12083} \BibitemShut {NoStop}%
\bibitem [{\citenamefont {Xu}\ \emph {et~al.}(2021)\citenamefont {Xu},
  \citenamefont {Al~Ezzi}, \citenamefont {Balakrishnan}, \citenamefont
  {Garcia-Ruiz}, \citenamefont {Tsim}, \citenamefont {Mullan}, \citenamefont
  {Barrier}, \citenamefont {Xin}, \citenamefont {Piot}, \citenamefont
  {Taniguchi}, \citenamefont {Watanabe}, \citenamefont {Carvalho},
  \citenamefont {Mishchenko}, \citenamefont {Geim}, \citenamefont {Fal'ko},
  \citenamefont {Adam}, \citenamefont {Neto}, \citenamefont {Novoselov},\ and\
  \citenamefont {Shi}}]{Xu_2021}%
  \BibitemOpen
  \bibfield  {author} {\bibinfo {author} {\bibfnamefont {S.}~\bibnamefont
  {Xu}}, \bibinfo {author} {\bibfnamefont {M.~M.}\ \bibnamefont {Al~Ezzi}},
  \bibinfo {author} {\bibfnamefont {N.}~\bibnamefont {Balakrishnan}}, \bibinfo
  {author} {\bibfnamefont {A.}~\bibnamefont {Garcia-Ruiz}}, \bibinfo {author}
  {\bibfnamefont {B.}~\bibnamefont {Tsim}}, \bibinfo {author} {\bibfnamefont
  {C.}~\bibnamefont {Mullan}}, \bibinfo {author} {\bibfnamefont
  {J.}~\bibnamefont {Barrier}}, \bibinfo {author} {\bibfnamefont
  {N.}~\bibnamefont {Xin}}, \bibinfo {author} {\bibfnamefont {B.~A.}\
  \bibnamefont {Piot}}, \bibinfo {author} {\bibfnamefont {T.}~\bibnamefont
  {Taniguchi}}, \bibinfo {author} {\bibfnamefont {K.}~\bibnamefont {Watanabe}},
  \bibinfo {author} {\bibfnamefont {A.}~\bibnamefont {Carvalho}}, \bibinfo
  {author} {\bibfnamefont {A.}~\bibnamefont {Mishchenko}}, \bibinfo {author}
  {\bibfnamefont {A.~K.}\ \bibnamefont {Geim}}, \bibinfo {author}
  {\bibfnamefont {V.~I.}\ \bibnamefont {Fal'ko}}, \bibinfo {author}
  {\bibfnamefont {S.}~\bibnamefont {Adam}}, \bibinfo {author} {\bibfnamefont
  {A.~H.~C.}\ \bibnamefont {Neto}}, \bibinfo {author} {\bibfnamefont {K.~S.}\
  \bibnamefont {Novoselov}}, \ and\ \bibinfo {author} {\bibfnamefont
  {Y.}~\bibnamefont {Shi}},\ }\href {\doibase 10.1038/s41567-021-01172-9}
  {\bibfield  {journal} {\bibinfo  {journal} {Nature Physics}\ }\textbf
  {\bibinfo {volume} {17}},\ \bibinfo {pages} {619} (\bibinfo {year}
  {2021})}\BibitemShut {NoStop}%
\bibitem [{\citenamefont {Choi}\ \emph {et~al.}(2021)\citenamefont {Choi},
  \citenamefont {Kim}, \citenamefont {Peng}, \citenamefont {Thomson},
  \citenamefont {Lewandowski}, \citenamefont {Polski}, \citenamefont {Zhang},
  \citenamefont {Arora}, \citenamefont {Watanabe}, \citenamefont {Taniguchi},
  \citenamefont {Alicea},\ and\ \citenamefont {Nadj-Perge}}]{Choi_2021}%
  \BibitemOpen
  \bibfield  {author} {\bibinfo {author} {\bibfnamefont {Y.}~\bibnamefont
  {Choi}}, \bibinfo {author} {\bibfnamefont {H.}~\bibnamefont {Kim}}, \bibinfo
  {author} {\bibfnamefont {Y.}~\bibnamefont {Peng}}, \bibinfo {author}
  {\bibfnamefont {A.}~\bibnamefont {Thomson}}, \bibinfo {author} {\bibfnamefont
  {C.}~\bibnamefont {Lewandowski}}, \bibinfo {author} {\bibfnamefont
  {R.}~\bibnamefont {Polski}}, \bibinfo {author} {\bibfnamefont
  {Y.}~\bibnamefont {Zhang}}, \bibinfo {author} {\bibfnamefont {H.~S.}\
  \bibnamefont {Arora}}, \bibinfo {author} {\bibfnamefont {K.}~\bibnamefont
  {Watanabe}}, \bibinfo {author} {\bibfnamefont {T.}~\bibnamefont {Taniguchi}},
  \bibinfo {author} {\bibfnamefont {J.}~\bibnamefont {Alicea}}, \ and\ \bibinfo
  {author} {\bibfnamefont {S.}~\bibnamefont {Nadj-Perge}},\ }\href {\doibase
  10.1038/s41586-020-03159-7} {\bibfield  {journal} {\bibinfo  {journal}
  {Nature}\ }\textbf {\bibinfo {volume} {589}},\ \bibinfo {pages} {536}
  (\bibinfo {year} {2021})}\BibitemShut {NoStop}%
\bibitem [{\citenamefont {Park}\ \emph {et~al.}(2022)\citenamefont {Park},
  \citenamefont {Cao}, \citenamefont {Xia}, \citenamefont {Sun}, \citenamefont
  {Watanabe}, \citenamefont {Taniguchi},\ and\ \citenamefont
  {Jarillo-Herrero}}]{Park2022}%
  \BibitemOpen
  \bibfield  {author} {\bibinfo {author} {\bibfnamefont {J.~M.}\ \bibnamefont
  {Park}}, \bibinfo {author} {\bibfnamefont {Y.}~\bibnamefont {Cao}}, \bibinfo
  {author} {\bibfnamefont {L.-Q.}\ \bibnamefont {Xia}}, \bibinfo {author}
  {\bibfnamefont {S.}~\bibnamefont {Sun}}, \bibinfo {author} {\bibfnamefont
  {K.}~\bibnamefont {Watanabe}}, \bibinfo {author} {\bibfnamefont
  {T.}~\bibnamefont {Taniguchi}}, \ and\ \bibinfo {author} {\bibfnamefont
  {P.}~\bibnamefont {Jarillo-Herrero}},\ }\href {\doibase
  10.1038/s41563-022-01287-1} {\bibfield  {journal} {\bibinfo  {journal}
  {Nature Materials}\ }\textbf {\bibinfo {volume} {21}},\ \bibinfo {pages}
  {877} (\bibinfo {year} {2022})}\BibitemShut {NoStop}%
\bibitem [{\citenamefont {Uri}\ \emph {et~al.}(2023)\citenamefont {Uri},
  \citenamefont {de~la Barrera}, \citenamefont {Randeria}, \citenamefont
  {Rodan-Legrain}, \citenamefont {Devakul}, \citenamefont {Crowley},
  \citenamefont {Paul}, \citenamefont {Watanabe}, \citenamefont {Taniguchi},
  \citenamefont {Lifshitz}, \citenamefont {Fu}, \citenamefont {Ashoori},\ and\
  \citenamefont {Jarillo-Herrero}}]{uri2023superconductivity}%
  \BibitemOpen
  \bibfield  {author} {\bibinfo {author} {\bibfnamefont {A.}~\bibnamefont
  {Uri}}, \bibinfo {author} {\bibfnamefont {S.~C.}\ \bibnamefont {de~la
  Barrera}}, \bibinfo {author} {\bibfnamefont {M.~T.}\ \bibnamefont
  {Randeria}}, \bibinfo {author} {\bibfnamefont {D.}~\bibnamefont
  {Rodan-Legrain}}, \bibinfo {author} {\bibfnamefont {T.}~\bibnamefont
  {Devakul}}, \bibinfo {author} {\bibfnamefont {P.~J.~D.}\ \bibnamefont
  {Crowley}}, \bibinfo {author} {\bibfnamefont {N.}~\bibnamefont {Paul}},
  \bibinfo {author} {\bibfnamefont {K.}~\bibnamefont {Watanabe}}, \bibinfo
  {author} {\bibfnamefont {T.}~\bibnamefont {Taniguchi}}, \bibinfo {author}
  {\bibfnamefont {R.}~\bibnamefont {Lifshitz}}, \bibinfo {author}
  {\bibfnamefont {L.}~\bibnamefont {Fu}}, \bibinfo {author} {\bibfnamefont
  {R.~C.}\ \bibnamefont {Ashoori}}, \ and\ \bibinfo {author} {\bibfnamefont
  {P.}~\bibnamefont {Jarillo-Herrero}},\ }\href@noop {} {\enquote {\bibinfo
  {title} {Superconductivity and strong interactions in a tunable moir\'e
  quasiperiodic crystal},}\ } (\bibinfo {year} {2023}),\ \Eprint
  {http://arxiv.org/abs/2302.00686} {arXiv:2302.00686 [cond-mat.mes-hall]}
  \BibitemShut {NoStop}%
\bibitem [{\citenamefont {Naik}\ and\ \citenamefont {Jain}(2018)}]{Naik_2018}%
  \BibitemOpen
  \bibfield  {author} {\bibinfo {author} {\bibfnamefont {M.~H.}\ \bibnamefont
  {Naik}}\ and\ \bibinfo {author} {\bibfnamefont {M.}~\bibnamefont {Jain}},\
  }\href {\doibase 10.1103/physrevlett.121.266401} {\bibfield  {journal}
  {\bibinfo  {journal} {Physical Review Letters}\ }\textbf {\bibinfo {volume}
  {121}},\ \bibinfo {pages} {266401} (\bibinfo {year} {2018})}\BibitemShut
  {NoStop}%
\bibitem [{\citenamefont {Tang}\ \emph {et~al.}(2020)\citenamefont {Tang},
  \citenamefont {Li}, \citenamefont {Li}, \citenamefont {Xu}, \citenamefont
  {Liu}, \citenamefont {Barmak}, \citenamefont {Watanabe}, \citenamefont
  {Taniguchi}, \citenamefont {MacDonald}, \citenamefont {Shan},\ and\
  \citenamefont {Mak}}]{Tang2020}%
  \BibitemOpen
  \bibfield  {author} {\bibinfo {author} {\bibfnamefont {Y.}~\bibnamefont
  {Tang}}, \bibinfo {author} {\bibfnamefont {L.}~\bibnamefont {Li}}, \bibinfo
  {author} {\bibfnamefont {T.}~\bibnamefont {Li}}, \bibinfo {author}
  {\bibfnamefont {Y.}~\bibnamefont {Xu}}, \bibinfo {author} {\bibfnamefont
  {S.}~\bibnamefont {Liu}}, \bibinfo {author} {\bibfnamefont {K.}~\bibnamefont
  {Barmak}}, \bibinfo {author} {\bibfnamefont {K.}~\bibnamefont {Watanabe}},
  \bibinfo {author} {\bibfnamefont {T.}~\bibnamefont {Taniguchi}}, \bibinfo
  {author} {\bibfnamefont {A.~H.}\ \bibnamefont {MacDonald}}, \bibinfo {author}
  {\bibfnamefont {J.}~\bibnamefont {Shan}}, \ and\ \bibinfo {author}
  {\bibfnamefont {K.~F.}\ \bibnamefont {Mak}},\ }\href@noop {} {\bibfield
  {journal} {\bibinfo  {journal} {Nature}\ }\textbf {\bibinfo {volume} {579}},\
  \bibinfo {pages} {353} (\bibinfo {year} {2020})}\BibitemShut {NoStop}%
\bibitem [{\citenamefont {Wang}\ \emph {et~al.}(2020)\citenamefont {Wang},
  \citenamefont {Shih}, \citenamefont {Ghiotto}, \citenamefont {Xian},
  \citenamefont {Rhodes}, \citenamefont {Tan}, \citenamefont {Claassen},
  \citenamefont {Kennes}, \citenamefont {Bai}, \citenamefont {Kim},
  \citenamefont {Watanabe}, \citenamefont {Taniguchi}, \citenamefont {Zhu},
  \citenamefont {Hone}, \citenamefont {Rubio}, \citenamefont {Pasupathy},\ and\
  \citenamefont {Dean}}]{Wang2020}%
  \BibitemOpen
  \bibfield  {author} {\bibinfo {author} {\bibfnamefont {L.}~\bibnamefont
  {Wang}}, \bibinfo {author} {\bibfnamefont {E.-M.}\ \bibnamefont {Shih}},
  \bibinfo {author} {\bibfnamefont {A.}~\bibnamefont {Ghiotto}}, \bibinfo
  {author} {\bibfnamefont {L.}~\bibnamefont {Xian}}, \bibinfo {author}
  {\bibfnamefont {D.~A.}\ \bibnamefont {Rhodes}}, \bibinfo {author}
  {\bibfnamefont {C.}~\bibnamefont {Tan}}, \bibinfo {author} {\bibfnamefont
  {M.}~\bibnamefont {Claassen}}, \bibinfo {author} {\bibfnamefont {D.~M.}\
  \bibnamefont {Kennes}}, \bibinfo {author} {\bibfnamefont {Y.}~\bibnamefont
  {Bai}}, \bibinfo {author} {\bibfnamefont {B.}~\bibnamefont {Kim}}, \bibinfo
  {author} {\bibfnamefont {K.}~\bibnamefont {Watanabe}}, \bibinfo {author}
  {\bibfnamefont {T.}~\bibnamefont {Taniguchi}}, \bibinfo {author}
  {\bibfnamefont {X.}~\bibnamefont {Zhu}}, \bibinfo {author} {\bibfnamefont
  {J.}~\bibnamefont {Hone}}, \bibinfo {author} {\bibfnamefont {A.}~\bibnamefont
  {Rubio}}, \bibinfo {author} {\bibfnamefont {A.~N.}\ \bibnamefont
  {Pasupathy}}, \ and\ \bibinfo {author} {\bibfnamefont {C.~R.}\ \bibnamefont
  {Dean}},\ }\href {\doibase 10.1038/s41563-020-0708-6} {\bibfield  {journal}
  {\bibinfo  {journal} {Nature Materials}\ }\textbf {\bibinfo {volume} {19}},\
  \bibinfo {pages} {861} (\bibinfo {year} {2020})}\BibitemShut {NoStop}%
\bibitem [{\citenamefont {Zhang}\ \emph {et~al.}(2020)\citenamefont {Zhang},
  \citenamefont {Zhang}, \citenamefont {Wu}, \citenamefont {Wang},
  \citenamefont {Gogna}, \citenamefont {Hou}, \citenamefont {Watanabe},
  \citenamefont {Taniguchi}, \citenamefont {Kulkarni}, \citenamefont {Kuo},
  \citenamefont {Forrest},\ and\ \citenamefont {Deng}}]{Zhang_2020}%
  \BibitemOpen
  \bibfield  {author} {\bibinfo {author} {\bibfnamefont {L.}~\bibnamefont
  {Zhang}}, \bibinfo {author} {\bibfnamefont {Z.}~\bibnamefont {Zhang}},
  \bibinfo {author} {\bibfnamefont {F.}~\bibnamefont {Wu}}, \bibinfo {author}
  {\bibfnamefont {D.}~\bibnamefont {Wang}}, \bibinfo {author} {\bibfnamefont
  {R.}~\bibnamefont {Gogna}}, \bibinfo {author} {\bibfnamefont
  {S.}~\bibnamefont {Hou}}, \bibinfo {author} {\bibfnamefont {K.}~\bibnamefont
  {Watanabe}}, \bibinfo {author} {\bibfnamefont {T.}~\bibnamefont {Taniguchi}},
  \bibinfo {author} {\bibfnamefont {K.}~\bibnamefont {Kulkarni}}, \bibinfo
  {author} {\bibfnamefont {T.}~\bibnamefont {Kuo}}, \bibinfo {author}
  {\bibfnamefont {S.~R.}\ \bibnamefont {Forrest}}, \ and\ \bibinfo {author}
  {\bibfnamefont {H.}~\bibnamefont {Deng}},\ }\href {\doibase
  10.1038/s41467-020-19466-6} {\bibfield  {journal} {\bibinfo  {journal}
  {Nature Communications}\ }\textbf {\bibinfo {volume} {11}},\ \bibinfo {pages}
  {5888} (\bibinfo {year} {2020})}\BibitemShut {NoStop}%
\bibitem [{\citenamefont {Shabani}\ \emph {et~al.}(2021)\citenamefont
  {Shabani}, \citenamefont {Halbertal}, \citenamefont {Wu}, \citenamefont
  {Chen}, \citenamefont {Liu}, \citenamefont {Hone}, \citenamefont {Yao},
  \citenamefont {Basov}, \citenamefont {Zhu},\ and\ \citenamefont
  {Pasupathy}}]{Shabani_2021}%
  \BibitemOpen
  \bibfield  {author} {\bibinfo {author} {\bibfnamefont {S.}~\bibnamefont
  {Shabani}}, \bibinfo {author} {\bibfnamefont {D.}~\bibnamefont {Halbertal}},
  \bibinfo {author} {\bibfnamefont {W.}~\bibnamefont {Wu}}, \bibinfo {author}
  {\bibfnamefont {M.}~\bibnamefont {Chen}}, \bibinfo {author} {\bibfnamefont
  {S.}~\bibnamefont {Liu}}, \bibinfo {author} {\bibfnamefont {J.}~\bibnamefont
  {Hone}}, \bibinfo {author} {\bibfnamefont {W.}~\bibnamefont {Yao}}, \bibinfo
  {author} {\bibfnamefont {D.~N.}\ \bibnamefont {Basov}}, \bibinfo {author}
  {\bibfnamefont {X.}~\bibnamefont {Zhu}}, \ and\ \bibinfo {author}
  {\bibfnamefont {A.~N.}\ \bibnamefont {Pasupathy}},\ }\href {\doibase
  10.1038/s41567-021-01174-7} {\bibfield  {journal} {\bibinfo  {journal}
  {Nature Physics}\ }\textbf {\bibinfo {volume} {17}},\ \bibinfo {pages} {720}
  (\bibinfo {year} {2021})}\BibitemShut {NoStop}%
\bibitem [{\citenamefont {Xu}\ \emph {et~al.}(2022)\citenamefont {Xu},
  \citenamefont {Kang}, \citenamefont {Watanabe}, \citenamefont {Taniguchi},
  \citenamefont {Mak},\ and\ \citenamefont {Shan}}]{Xu2022}%
  \BibitemOpen
  \bibfield  {author} {\bibinfo {author} {\bibfnamefont {Y.}~\bibnamefont
  {Xu}}, \bibinfo {author} {\bibfnamefont {K.}~\bibnamefont {Kang}}, \bibinfo
  {author} {\bibfnamefont {K.}~\bibnamefont {Watanabe}}, \bibinfo {author}
  {\bibfnamefont {T.}~\bibnamefont {Taniguchi}}, \bibinfo {author}
  {\bibfnamefont {K.~F.}\ \bibnamefont {Mak}}, \ and\ \bibinfo {author}
  {\bibfnamefont {J.}~\bibnamefont {Shan}},\ }\href {\doibase
  10.1038/s41565-022-01180-7} {\bibfield  {journal} {\bibinfo  {journal}
  {Nature Nanotechnology}\ }\textbf {\bibinfo {volume} {17}},\ \bibinfo {pages}
  {934} (\bibinfo {year} {2022})}\BibitemShut {NoStop}%
\bibitem [{\citenamefont {Xiong}\ \emph {et~al.}(2022)\citenamefont {Xiong},
  \citenamefont {Wang}, \citenamefont {Zhu}, \citenamefont {Xu}, \citenamefont
  {Wu}, \citenamefont {Chen}, \citenamefont {Ma}, \citenamefont {Liu},
  \citenamefont {Chen}, \citenamefont {Watanabe}, \citenamefont {Taniguchi},
  \citenamefont {Shi}, \citenamefont {Chen}, \citenamefont {Lu}, \citenamefont
  {Zhan}, \citenamefont {Hao},\ and\ \citenamefont {Xu}}]{Xiong2022}%
  \BibitemOpen
  \bibfield  {author} {\bibinfo {author} {\bibfnamefont {Y.}~\bibnamefont
  {Xiong}}, \bibinfo {author} {\bibfnamefont {Y.}~\bibnamefont {Wang}},
  \bibinfo {author} {\bibfnamefont {R.}~\bibnamefont {Zhu}}, \bibinfo {author}
  {\bibfnamefont {H.}~\bibnamefont {Xu}}, \bibinfo {author} {\bibfnamefont
  {C.}~\bibnamefont {Wu}}, \bibinfo {author} {\bibfnamefont {J.}~\bibnamefont
  {Chen}}, \bibinfo {author} {\bibfnamefont {Y.}~\bibnamefont {Ma}}, \bibinfo
  {author} {\bibfnamefont {Y.}~\bibnamefont {Liu}}, \bibinfo {author}
  {\bibfnamefont {Y.}~\bibnamefont {Chen}}, \bibinfo {author} {\bibfnamefont
  {K.}~\bibnamefont {Watanabe}}, \bibinfo {author} {\bibfnamefont
  {T.}~\bibnamefont {Taniguchi}}, \bibinfo {author} {\bibfnamefont
  {M.}~\bibnamefont {Shi}}, \bibinfo {author} {\bibfnamefont {X.}~\bibnamefont
  {Chen}}, \bibinfo {author} {\bibfnamefont {Y.}~\bibnamefont {Lu}}, \bibinfo
  {author} {\bibfnamefont {P.}~\bibnamefont {Zhan}}, \bibinfo {author}
  {\bibfnamefont {Y.}~\bibnamefont {Hao}}, \ and\ \bibinfo {author}
  {\bibfnamefont {F.}~\bibnamefont {Xu}},\ }\href {\doibase
  10.1126/sciadv.abo0375} {\bibfield  {journal} {\bibinfo  {journal} {Science
  Advances}\ }\textbf {\bibinfo {volume} {8}},\ \bibinfo {pages} {eabo0375}
  (\bibinfo {year} {2022})},\ \Eprint
  {http://arxiv.org/abs/https://www.science.org/doi/pdf/10.1126/sciadv.abo0375}
  {https://www.science.org/doi/pdf/10.1126/sciadv.abo0375} \BibitemShut
  {NoStop}%
\bibitem [{\citenamefont {Meng}\ \emph {et~al.}(2023)\citenamefont {Meng},
  \citenamefont {Wang}, \citenamefont {Han}, \citenamefont {Liu}, \citenamefont
  {Wen}, \citenamefont {Gao}, \citenamefont {Wang}, \citenamefont {Chin},\ and\
  \citenamefont {Zhang}}]{Meng2023}%
  \BibitemOpen
  \bibfield  {author} {\bibinfo {author} {\bibfnamefont {Z.}~\bibnamefont
  {Meng}}, \bibinfo {author} {\bibfnamefont {L.}~\bibnamefont {Wang}}, \bibinfo
  {author} {\bibfnamefont {W.}~\bibnamefont {Han}}, \bibinfo {author}
  {\bibfnamefont {F.}~\bibnamefont {Liu}}, \bibinfo {author} {\bibfnamefont
  {K.}~\bibnamefont {Wen}}, \bibinfo {author} {\bibfnamefont {C.}~\bibnamefont
  {Gao}}, \bibinfo {author} {\bibfnamefont {P.}~\bibnamefont {Wang}}, \bibinfo
  {author} {\bibfnamefont {C.}~\bibnamefont {Chin}}, \ and\ \bibinfo {author}
  {\bibfnamefont {J.}~\bibnamefont {Zhang}},\ }\href {\doibase
  10.1038/s41586-023-05695-4} {\bibfield  {journal} {\bibinfo  {journal}
  {Nature}\ }\textbf {\bibinfo {volume} {615}},\ \bibinfo {pages} {231}
  (\bibinfo {year} {2023})}\BibitemShut {NoStop}%
\bibitem [{\citenamefont {Kennes}\ \emph {et~al.}(2021)\citenamefont {Kennes},
  \citenamefont {Claassen}, \citenamefont {Xian}, \citenamefont {Georges},
  \citenamefont {Millis}, \citenamefont {Hone}, \citenamefont {Dean},
  \citenamefont {Basov}, \citenamefont {Pasupathy},\ and\ \citenamefont
  {Rubio}}]{Kennes2021}%
  \BibitemOpen
  \bibfield  {author} {\bibinfo {author} {\bibfnamefont {D.~M.}\ \bibnamefont
  {Kennes}}, \bibinfo {author} {\bibfnamefont {M.}~\bibnamefont {Claassen}},
  \bibinfo {author} {\bibfnamefont {L.}~\bibnamefont {Xian}}, \bibinfo {author}
  {\bibfnamefont {A.}~\bibnamefont {Georges}}, \bibinfo {author} {\bibfnamefont
  {A.~J.}\ \bibnamefont {Millis}}, \bibinfo {author} {\bibfnamefont
  {J.}~\bibnamefont {Hone}}, \bibinfo {author} {\bibfnamefont {C.~R.}\
  \bibnamefont {Dean}}, \bibinfo {author} {\bibfnamefont {D.~N.}\ \bibnamefont
  {Basov}}, \bibinfo {author} {\bibfnamefont {A.~N.}\ \bibnamefont
  {Pasupathy}}, \ and\ \bibinfo {author} {\bibfnamefont {A.}~\bibnamefont
  {Rubio}},\ }\href {\doibase 10.1038/s41567-020-01154-3} {\bibfield  {journal}
  {\bibinfo  {journal} {Nature Physics}\ }\textbf {\bibinfo {volume} {17}},\
  \bibinfo {pages} {155} (\bibinfo {year} {2021})}\BibitemShut {NoStop}%
\bibitem [{\citenamefont {Xu}\ and\ \citenamefont {Balents}(2018)}]{Xu_2018}%
  \BibitemOpen
  \bibfield  {author} {\bibinfo {author} {\bibfnamefont {C.}~\bibnamefont
  {Xu}}\ and\ \bibinfo {author} {\bibfnamefont {L.}~\bibnamefont {Balents}},\
  }\href {\doibase 10.1103/physrevlett.121.087001} {\bibfield  {journal}
  {\bibinfo  {journal} {Physical Review Letters}\ }\textbf {\bibinfo {volume}
  {121}},\ \bibinfo {pages} {087001} (\bibinfo {year} {2018})}\BibitemShut
  {NoStop}%
\bibitem [{\citenamefont {Po}\ \emph {et~al.}(2018)\citenamefont {Po},
  \citenamefont {Zou}, \citenamefont {Vishwanath},\ and\ \citenamefont
  {Senthil}}]{Po_2018}%
  \BibitemOpen
  \bibfield  {author} {\bibinfo {author} {\bibfnamefont {H.~C.}\ \bibnamefont
  {Po}}, \bibinfo {author} {\bibfnamefont {L.}~\bibnamefont {Zou}}, \bibinfo
  {author} {\bibfnamefont {A.}~\bibnamefont {Vishwanath}}, \ and\ \bibinfo
  {author} {\bibfnamefont {T.}~\bibnamefont {Senthil}},\ }\href {\doibase
  10.1103/physrevx.8.031089} {\bibfield  {journal} {\bibinfo  {journal}
  {Physical Review X}\ }\textbf {\bibinfo {volume} {8}},\ \bibinfo {pages}
  {031089} (\bibinfo {year} {2018})}\BibitemShut {NoStop}%
\bibitem [{\citenamefont {Wu}\ \emph {et~al.}(2018{\natexlab{a}})\citenamefont
  {Wu}, \citenamefont {Lovorn}, \citenamefont {Tutuc},\ and\ \citenamefont
  {MacDonald}}]{Fengcheng_2018}%
  \BibitemOpen
  \bibfield  {author} {\bibinfo {author} {\bibfnamefont {F.}~\bibnamefont
  {Wu}}, \bibinfo {author} {\bibfnamefont {T.}~\bibnamefont {Lovorn}}, \bibinfo
  {author} {\bibfnamefont {E.}~\bibnamefont {Tutuc}}, \ and\ \bibinfo {author}
  {\bibfnamefont {A.~H.}\ \bibnamefont {MacDonald}},\ }\href {\doibase
  10.1103/PhysRevLett.121.026402} {\bibfield  {journal} {\bibinfo  {journal}
  {Phys. Rev. Lett.}\ }\textbf {\bibinfo {volume} {121}},\ \bibinfo {pages}
  {026402} (\bibinfo {year} {2018}{\natexlab{a}})}\BibitemShut {NoStop}%
\bibitem [{\citenamefont {Kang}\ and\ \citenamefont {Vafek}(2018)}]{Kang_2018}%
  \BibitemOpen
  \bibfield  {author} {\bibinfo {author} {\bibfnamefont {J.}~\bibnamefont
  {Kang}}\ and\ \bibinfo {author} {\bibfnamefont {O.}~\bibnamefont {Vafek}},\
  }\href {\doibase 10.1103/physrevx.8.031088} {\bibfield  {journal} {\bibinfo
  {journal} {Physical Review X}\ }\textbf {\bibinfo {volume} {8}},\ \bibinfo
  {pages} {031088} (\bibinfo {year} {2018})}\BibitemShut {NoStop}%
\bibitem [{\citenamefont {Kang}\ and\ \citenamefont {Vafek}(2019)}]{Kang_2019}%
  \BibitemOpen
  \bibfield  {author} {\bibinfo {author} {\bibfnamefont {J.}~\bibnamefont
  {Kang}}\ and\ \bibinfo {author} {\bibfnamefont {O.}~\bibnamefont {Vafek}},\
  }\href {\doibase 10.1103/physrevlett.122.246401} {\bibfield  {journal}
  {\bibinfo  {journal} {Physical Review Letters}\ }\textbf {\bibinfo {volume}
  {122}},\ \bibinfo {pages} {246401} (\bibinfo {year} {2019})}\BibitemShut
  {NoStop}%
\bibitem [{\citenamefont {Koshino}(2019)}]{Koshino_2019}%
  \BibitemOpen
  \bibfield  {author} {\bibinfo {author} {\bibfnamefont {M.}~\bibnamefont
  {Koshino}},\ }\href {\doibase 10.1103/physrevb.99.235406} {\bibfield
  {journal} {\bibinfo  {journal} {Physical Review B}\ }\textbf {\bibinfo
  {volume} {99}},\ \bibinfo {pages} {235406} (\bibinfo {year}
  {2019})}\BibitemShut {NoStop}%
\bibitem [{\citenamefont {Hejazi}\ \emph {et~al.}(2019)\citenamefont {Hejazi},
  \citenamefont {Liu}, \citenamefont {Shapourian}, \citenamefont {Chen},\ and\
  \citenamefont {Balents}}]{Hejazi_2019}%
  \BibitemOpen
  \bibfield  {author} {\bibinfo {author} {\bibfnamefont {K.}~\bibnamefont
  {Hejazi}}, \bibinfo {author} {\bibfnamefont {C.}~\bibnamefont {Liu}},
  \bibinfo {author} {\bibfnamefont {H.}~\bibnamefont {Shapourian}}, \bibinfo
  {author} {\bibfnamefont {X.}~\bibnamefont {Chen}}, \ and\ \bibinfo {author}
  {\bibfnamefont {L.}~\bibnamefont {Balents}},\ }\href {\doibase
  10.1103/PhysRevB.99.035111} {\bibfield  {journal} {\bibinfo  {journal} {Phys.
  Rev. B}\ }\textbf {\bibinfo {volume} {99}},\ \bibinfo {pages} {035111}
  (\bibinfo {year} {2019})}\BibitemShut {NoStop}%
\bibitem [{\citenamefont {Chebrolu}\ \emph {et~al.}(2019)\citenamefont
  {Chebrolu}, \citenamefont {Chittari},\ and\ \citenamefont
  {Jung}}]{Chebrolu_2019}%
  \BibitemOpen
  \bibfield  {author} {\bibinfo {author} {\bibfnamefont {N.~R.}\ \bibnamefont
  {Chebrolu}}, \bibinfo {author} {\bibfnamefont {B.~L.}\ \bibnamefont
  {Chittari}}, \ and\ \bibinfo {author} {\bibfnamefont {J.}~\bibnamefont
  {Jung}},\ }\href {\doibase 10.1103/PhysRevB.99.235417} {\bibfield  {journal}
  {\bibinfo  {journal} {Phys. Rev. B}\ }\textbf {\bibinfo {volume} {99}},\
  \bibinfo {pages} {235417} (\bibinfo {year} {2019})}\BibitemShut {NoStop}%
\bibitem [{\citenamefont {Cea}\ \emph {et~al.}(2019)\citenamefont {Cea},
  \citenamefont {Walet},\ and\ \citenamefont {Guinea}}]{Cea_2019}%
  \BibitemOpen
  \bibfield  {author} {\bibinfo {author} {\bibfnamefont {T.}~\bibnamefont
  {Cea}}, \bibinfo {author} {\bibfnamefont {N.~R.}\ \bibnamefont {Walet}}, \
  and\ \bibinfo {author} {\bibfnamefont {F.}~\bibnamefont {Guinea}},\ }\href
  {\doibase 10.1103/PhysRevB.100.205113} {\bibfield  {journal} {\bibinfo
  {journal} {Phys. Rev. B}\ }\textbf {\bibinfo {volume} {100}},\ \bibinfo
  {pages} {205113} (\bibinfo {year} {2019})}\BibitemShut {NoStop}%
\bibitem [{\citenamefont {Liu}\ \emph {et~al.}(2019)\citenamefont {Liu},
  \citenamefont {Ma}, \citenamefont {Gao},\ and\ \citenamefont
  {Dai}}]{Liu_2019}%
  \BibitemOpen
  \bibfield  {author} {\bibinfo {author} {\bibfnamefont {J.}~\bibnamefont
  {Liu}}, \bibinfo {author} {\bibfnamefont {Z.}~\bibnamefont {Ma}}, \bibinfo
  {author} {\bibfnamefont {J.}~\bibnamefont {Gao}}, \ and\ \bibinfo {author}
  {\bibfnamefont {X.}~\bibnamefont {Dai}},\ }\href {\doibase
  10.1103/PhysRevX.9.031021} {\bibfield  {journal} {\bibinfo  {journal} {Phys.
  Rev. X}\ }\textbf {\bibinfo {volume} {9}},\ \bibinfo {pages} {031021}
  (\bibinfo {year} {2019})}\BibitemShut {NoStop}%
\bibitem [{\citenamefont {Liang}\ \emph {et~al.}(2020)\citenamefont {Liang},
  \citenamefont {Goodwin}, \citenamefont {Vitale}, \citenamefont {Corsetti},
  \citenamefont {Mostofi},\ and\ \citenamefont {Lischner}}]{Liang_2020}%
  \BibitemOpen
  \bibfield  {author} {\bibinfo {author} {\bibfnamefont {X.}~\bibnamefont
  {Liang}}, \bibinfo {author} {\bibfnamefont {Z.~A.~H.}\ \bibnamefont
  {Goodwin}}, \bibinfo {author} {\bibfnamefont {V.}~\bibnamefont {Vitale}},
  \bibinfo {author} {\bibfnamefont {F.}~\bibnamefont {Corsetti}}, \bibinfo
  {author} {\bibfnamefont {A.~A.}\ \bibnamefont {Mostofi}}, \ and\ \bibinfo
  {author} {\bibfnamefont {J.}~\bibnamefont {Lischner}},\ }\href {\doibase
  10.1103/physrevb.102.155146} {\bibfield  {journal} {\bibinfo  {journal}
  {Physical Review B}\ }\textbf {\bibinfo {volume} {102}},\ \bibinfo {pages}
  {155146} (\bibinfo {year} {2020})}\BibitemShut {NoStop}%
\bibitem [{\citenamefont {Tran}\ \emph {et~al.}(2020)\citenamefont {Tran},
  \citenamefont {Choi},\ and\ \citenamefont {Singh}}]{Tran_2020}%
  \BibitemOpen
  \bibfield  {author} {\bibinfo {author} {\bibfnamefont {K.}~\bibnamefont
  {Tran}}, \bibinfo {author} {\bibfnamefont {J.}~\bibnamefont {Choi}}, \ and\
  \bibinfo {author} {\bibfnamefont {A.}~\bibnamefont {Singh}},\ }\href
  {\doibase 10.1088/2053-1583/abd3e7} {\bibfield  {journal} {\bibinfo
  {journal} {2D Materials}\ }\textbf {\bibinfo {volume} {8}},\ \bibinfo {pages}
  {022002} (\bibinfo {year} {2020})}\BibitemShut {NoStop}%
\bibitem [{\citenamefont {Tritsaris}\ \emph {et~al.}(2020)\citenamefont
  {Tritsaris}, \citenamefont {Carr}, \citenamefont {Zhu}, \citenamefont {Xie},
  \citenamefont {Torrisi}, \citenamefont {Tang}, \citenamefont {Mattheakis},
  \citenamefont {Larson},\ and\ \citenamefont {Kaxiras}}]{Tritsaris_2020}%
  \BibitemOpen
  \bibfield  {author} {\bibinfo {author} {\bibfnamefont {G.~A.}\ \bibnamefont
  {Tritsaris}}, \bibinfo {author} {\bibfnamefont {S.}~\bibnamefont {Carr}},
  \bibinfo {author} {\bibfnamefont {Z.}~\bibnamefont {Zhu}}, \bibinfo {author}
  {\bibfnamefont {Y.}~\bibnamefont {Xie}}, \bibinfo {author} {\bibfnamefont
  {S.~B.}\ \bibnamefont {Torrisi}}, \bibinfo {author} {\bibfnamefont
  {J.}~\bibnamefont {Tang}}, \bibinfo {author} {\bibfnamefont {M.}~\bibnamefont
  {Mattheakis}}, \bibinfo {author} {\bibfnamefont {D.~T.}\ \bibnamefont
  {Larson}}, \ and\ \bibinfo {author} {\bibfnamefont {E.}~\bibnamefont
  {Kaxiras}},\ }\href {\doibase 10.1088/2053-1583/ab8f62} {\bibfield  {journal}
  {\bibinfo  {journal} {2D Materials}\ }\textbf {\bibinfo {volume} {7}},\
  \bibinfo {pages} {035028} (\bibinfo {year} {2020})}\BibitemShut {NoStop}%
\bibitem [{\citenamefont {Ramires}\ and\ \citenamefont
  {Lado}(2021)}]{Ramires2021}%
  \BibitemOpen
  \bibfield  {author} {\bibinfo {author} {\bibfnamefont {A.}~\bibnamefont
  {Ramires}}\ and\ \bibinfo {author} {\bibfnamefont {J.~L.}\ \bibnamefont
  {Lado}},\ }\href {\doibase 10.1103/PhysRevLett.127.026401} {\bibfield
  {journal} {\bibinfo  {journal} {Phys. Rev. Lett.}\ }\textbf {\bibinfo
  {volume} {127}},\ \bibinfo {pages} {026401} (\bibinfo {year}
  {2021})}\BibitemShut {NoStop}%
\bibitem [{\citenamefont {Lake}\ and\ \citenamefont
  {Senthil}(2021)}]{Lake2021}%
  \BibitemOpen
  \bibfield  {author} {\bibinfo {author} {\bibfnamefont {E.}~\bibnamefont
  {Lake}}\ and\ \bibinfo {author} {\bibfnamefont {T.}~\bibnamefont {Senthil}},\
  }\href {\doibase 10.1103/PhysRevB.104.174505} {\bibfield  {journal} {\bibinfo
   {journal} {Phys. Rev. B}\ }\textbf {\bibinfo {volume} {104}},\ \bibinfo
  {pages} {174505} (\bibinfo {year} {2021})}\BibitemShut {NoStop}%
\bibitem [{\citenamefont {Qin}\ and\ \citenamefont
  {MacDonald}(2021)}]{Qin2021}%
  \BibitemOpen
  \bibfield  {author} {\bibinfo {author} {\bibfnamefont {W.}~\bibnamefont
  {Qin}}\ and\ \bibinfo {author} {\bibfnamefont {A.~H.}\ \bibnamefont
  {MacDonald}},\ }\href {\doibase 10.1103/PhysRevLett.127.097001} {\bibfield
  {journal} {\bibinfo  {journal} {Phys. Rev. Lett.}\ }\textbf {\bibinfo
  {volume} {127}},\ \bibinfo {pages} {097001} (\bibinfo {year}
  {2021})}\BibitemShut {NoStop}%
\bibitem [{\citenamefont {Eaton}\ \emph {et~al.}(2022)\citenamefont {Eaton},
  \citenamefont {Li}, \citenamefont {Fertig},\ and\ \citenamefont
  {Seradjeh}}]{Eaton2022}%
  \BibitemOpen
  \bibfield  {author} {\bibinfo {author} {\bibfnamefont {A.}~\bibnamefont
  {Eaton}}, \bibinfo {author} {\bibfnamefont {Y.}~\bibnamefont {Li}}, \bibinfo
  {author} {\bibfnamefont {H.~A.}\ \bibnamefont {Fertig}}, \ and\ \bibinfo
  {author} {\bibfnamefont {B.}~\bibnamefont {Seradjeh}},\ }\href {\doibase
  10.1103/PhysRevB.106.045117} {\bibfield  {journal} {\bibinfo  {journal}
  {Phys. Rev. B}\ }\textbf {\bibinfo {volume} {106}},\ \bibinfo {pages}
  {045117} (\bibinfo {year} {2022})}\BibitemShut {NoStop}%
\bibitem [{\citenamefont {Zhou}\ \emph {et~al.}(2022)\citenamefont {Zhou},
  \citenamefont {Sheng},\ and\ \citenamefont {Kim}}]{Zhou2022}%
  \BibitemOpen
  \bibfield  {author} {\bibinfo {author} {\bibfnamefont {Y.}~\bibnamefont
  {Zhou}}, \bibinfo {author} {\bibfnamefont {D.~N.}\ \bibnamefont {Sheng}}, \
  and\ \bibinfo {author} {\bibfnamefont {E.-A.}\ \bibnamefont {Kim}},\ }\href
  {\doibase 10.1103/PhysRevLett.128.157602} {\bibfield  {journal} {\bibinfo
  {journal} {Phys. Rev. Lett.}\ }\textbf {\bibinfo {volume} {128}},\ \bibinfo
  {pages} {157602} (\bibinfo {year} {2022})}\BibitemShut {NoStop}%
\bibitem [{\citenamefont {Topp}\ \emph {et~al.}(2019)\citenamefont {Topp},
  \citenamefont {Jotzu}, \citenamefont {McIver}, \citenamefont {Xian},
  \citenamefont {Rubio},\ and\ \citenamefont {Sentef}}]{Topp_2019}%
  \BibitemOpen
  \bibfield  {author} {\bibinfo {author} {\bibfnamefont {G.~E.}\ \bibnamefont
  {Topp}}, \bibinfo {author} {\bibfnamefont {G.}~\bibnamefont {Jotzu}},
  \bibinfo {author} {\bibfnamefont {J.~W.}\ \bibnamefont {McIver}}, \bibinfo
  {author} {\bibfnamefont {L.}~\bibnamefont {Xian}}, \bibinfo {author}
  {\bibfnamefont {A.}~\bibnamefont {Rubio}}, \ and\ \bibinfo {author}
  {\bibfnamefont {M.~A.}\ \bibnamefont {Sentef}},\ }\href {\doibase
  10.1103/physrevresearch.1.023031} {\bibfield  {journal} {\bibinfo  {journal}
  {Physical Review Research}\ }\textbf {\bibinfo {volume} {1}},\ \bibinfo
  {pages} {023031} (\bibinfo {year} {2019})}\BibitemShut {NoStop}%
\bibitem [{\citenamefont {Li}\ \emph {et~al.}(2020)\citenamefont {Li},
  \citenamefont {Fertig},\ and\ \citenamefont {Seradjeh}}]{Li_2020}%
  \BibitemOpen
  \bibfield  {author} {\bibinfo {author} {\bibfnamefont {Y.}~\bibnamefont
  {Li}}, \bibinfo {author} {\bibfnamefont {H.~A.}\ \bibnamefont {Fertig}}, \
  and\ \bibinfo {author} {\bibfnamefont {B.}~\bibnamefont {Seradjeh}},\ }\href
  {\doibase 10.1103/physrevresearch.2.043275} {\bibfield  {journal} {\bibinfo
  {journal} {Physical Review Research}\ }\textbf {\bibinfo {volume} {2}},\
  \bibinfo {pages} {043275} (\bibinfo {year} {2020})}\BibitemShut {NoStop}%
\bibitem [{\citenamefont {Katz}\ \emph {et~al.}(2020)\citenamefont {Katz},
  \citenamefont {Refael},\ and\ \citenamefont {Lindner}}]{Katz_2020}%
  \BibitemOpen
  \bibfield  {author} {\bibinfo {author} {\bibfnamefont {O.}~\bibnamefont
  {Katz}}, \bibinfo {author} {\bibfnamefont {G.}~\bibnamefont {Refael}}, \ and\
  \bibinfo {author} {\bibfnamefont {N.~H.}\ \bibnamefont {Lindner}},\ }\href
  {\doibase 10.1103/physrevb.102.155123} {\bibfield  {journal} {\bibinfo
  {journal} {Physical Review B}\ }\textbf {\bibinfo {volume} {102}},\ \bibinfo
  {pages} {155123} (\bibinfo {year} {2020})}\BibitemShut {NoStop}%
\bibitem [{\citenamefont {Vogl}\ \emph {et~al.}(2020)\citenamefont {Vogl},
  \citenamefont {Rodriguez-Vega},\ and\ \citenamefont {Fiete}}]{Vogl2020}%
  \BibitemOpen
  \bibfield  {author} {\bibinfo {author} {\bibfnamefont {M.}~\bibnamefont
  {Vogl}}, \bibinfo {author} {\bibfnamefont {M.}~\bibnamefont
  {Rodriguez-Vega}}, \ and\ \bibinfo {author} {\bibfnamefont {G.~A.}\
  \bibnamefont {Fiete}},\ }\href {\doibase 10.1103/PhysRevB.101.235411}
  {\bibfield  {journal} {\bibinfo  {journal} {Phys. Rev. B}\ }\textbf {\bibinfo
  {volume} {101}},\ \bibinfo {pages} {235411} (\bibinfo {year}
  {2020})}\BibitemShut {NoStop}%
\bibitem [{\citenamefont {Vogl}\ \emph {et~al.}(2021)\citenamefont {Vogl},
  \citenamefont {Rodriguez-Vega}, \citenamefont {Flebus}, \citenamefont
  {MacDonald},\ and\ \citenamefont {Fiete}}]{Vogl2021}%
  \BibitemOpen
  \bibfield  {author} {\bibinfo {author} {\bibfnamefont {M.}~\bibnamefont
  {Vogl}}, \bibinfo {author} {\bibfnamefont {M.}~\bibnamefont
  {Rodriguez-Vega}}, \bibinfo {author} {\bibfnamefont {B.}~\bibnamefont
  {Flebus}}, \bibinfo {author} {\bibfnamefont {A.~H.}\ \bibnamefont
  {MacDonald}}, \ and\ \bibinfo {author} {\bibfnamefont {G.~A.}\ \bibnamefont
  {Fiete}},\ }\href {\doibase 10.1103/PhysRevB.103.014310} {\bibfield
  {journal} {\bibinfo  {journal} {Phys. Rev. B}\ }\textbf {\bibinfo {volume}
  {103}},\ \bibinfo {pages} {014310} (\bibinfo {year} {2021})}\BibitemShut
  {NoStop}%
\bibitem [{\citenamefont {Bistritzer}\ and\ \citenamefont
  {MacDonald}(2011)}]{Bistritzer_2011}%
  \BibitemOpen
  \bibfield  {author} {\bibinfo {author} {\bibfnamefont {R.}~\bibnamefont
  {Bistritzer}}\ and\ \bibinfo {author} {\bibfnamefont {A.~H.}\ \bibnamefont
  {MacDonald}},\ }\href {\doibase 10.1073/pnas.1108174108} {\bibfield
  {journal} {\bibinfo  {journal} {Proceedings of the National Academy of
  Sciences}\ }\textbf {\bibinfo {volume} {108}},\ \bibinfo {pages} {12233}
  (\bibinfo {year} {2011})}\BibitemShut {NoStop}%
\bibitem [{\citenamefont {Lopes~dos Santos}\ \emph {et~al.}(2007)\citenamefont
  {Lopes~dos Santos}, \citenamefont {Peres},\ and\ \citenamefont
  {Castro~Neto}}]{dosSantos_2007}%
  \BibitemOpen
  \bibfield  {author} {\bibinfo {author} {\bibfnamefont {J.~M.~B.}\
  \bibnamefont {Lopes~dos Santos}}, \bibinfo {author} {\bibfnamefont
  {N.~M.~R.}\ \bibnamefont {Peres}}, \ and\ \bibinfo {author} {\bibfnamefont
  {A.~H.}\ \bibnamefont {Castro~Neto}},\ }\href {\doibase
  10.1103/PhysRevLett.99.256802} {\bibfield  {journal} {\bibinfo  {journal}
  {Phys. Rev. Lett.}\ }\textbf {\bibinfo {volume} {99}},\ \bibinfo {pages}
  {256802} (\bibinfo {year} {2007})}\BibitemShut {NoStop}%
\bibitem [{\citenamefont {Shallcross}\ \emph {et~al.}(2008)\citenamefont
  {Shallcross}, \citenamefont {Sharma},\ and\ \citenamefont
  {Pankratov}}]{Shallcross_2008}%
  \BibitemOpen
  \bibfield  {author} {\bibinfo {author} {\bibfnamefont {S.}~\bibnamefont
  {Shallcross}}, \bibinfo {author} {\bibfnamefont {S.}~\bibnamefont {Sharma}},
  \ and\ \bibinfo {author} {\bibfnamefont {O.~A.}\ \bibnamefont {Pankratov}},\
  }\href {\doibase 10.1103/PhysRevLett.101.056803} {\bibfield  {journal}
  {\bibinfo  {journal} {Phys. Rev. Lett.}\ }\textbf {\bibinfo {volume} {101}},\
  \bibinfo {pages} {056803} (\bibinfo {year} {2008})}\BibitemShut {NoStop}%
\bibitem [{\citenamefont {Shallcross}\ \emph {et~al.}(2010)\citenamefont
  {Shallcross}, \citenamefont {Sharma}, \citenamefont {Kandelaki},\ and\
  \citenamefont {Pankratov}}]{Shallcross_2010}%
  \BibitemOpen
  \bibfield  {author} {\bibinfo {author} {\bibfnamefont {S.}~\bibnamefont
  {Shallcross}}, \bibinfo {author} {\bibfnamefont {S.}~\bibnamefont {Sharma}},
  \bibinfo {author} {\bibfnamefont {E.}~\bibnamefont {Kandelaki}}, \ and\
  \bibinfo {author} {\bibfnamefont {O.~A.}\ \bibnamefont {Pankratov}},\ }\href
  {\doibase 10.1103/PhysRevB.81.165105} {\bibfield  {journal} {\bibinfo
  {journal} {Phys. Rev. B}\ }\textbf {\bibinfo {volume} {81}},\ \bibinfo
  {pages} {165105} (\bibinfo {year} {2010})}\BibitemShut {NoStop}%
\bibitem [{\citenamefont {Mele}(2010)}]{Mele_2010}%
  \BibitemOpen
  \bibfield  {author} {\bibinfo {author} {\bibfnamefont {E.~J.}\ \bibnamefont
  {Mele}},\ }\href {\doibase 10.1103/PhysRevB.81.161405} {\bibfield  {journal}
  {\bibinfo  {journal} {Phys. Rev. B}\ }\textbf {\bibinfo {volume} {81}},\
  \bibinfo {pages} {161405} (\bibinfo {year} {2010})}\BibitemShut {NoStop}%
\bibitem [{\citenamefont {Mele}(2011)}]{Mele_2011}%
  \BibitemOpen
  \bibfield  {author} {\bibinfo {author} {\bibfnamefont {E.~J.}\ \bibnamefont
  {Mele}},\ }\href {\doibase 10.1103/PhysRevB.84.235439} {\bibfield  {journal}
  {\bibinfo  {journal} {Phys. Rev. B}\ }\textbf {\bibinfo {volume} {84}},\
  \bibinfo {pages} {235439} (\bibinfo {year} {2011})}\BibitemShut {NoStop}%
\bibitem [{\citenamefont {Carr}\ \emph {et~al.}(2017)\citenamefont {Carr},
  \citenamefont {Massatt}, \citenamefont {Fang}, \citenamefont {Cazeaux},
  \citenamefont {Luskin},\ and\ \citenamefont {Kaxiras}}]{Carr_2017}%
  \BibitemOpen
  \bibfield  {author} {\bibinfo {author} {\bibfnamefont {S.}~\bibnamefont
  {Carr}}, \bibinfo {author} {\bibfnamefont {D.}~\bibnamefont {Massatt}},
  \bibinfo {author} {\bibfnamefont {S.}~\bibnamefont {Fang}}, \bibinfo {author}
  {\bibfnamefont {P.}~\bibnamefont {Cazeaux}}, \bibinfo {author} {\bibfnamefont
  {M.}~\bibnamefont {Luskin}}, \ and\ \bibinfo {author} {\bibfnamefont
  {E.}~\bibnamefont {Kaxiras}},\ }\href {\doibase 10.1103/PhysRevB.95.075420}
  {\bibfield  {journal} {\bibinfo  {journal} {Phys. Rev. B}\ }\textbf {\bibinfo
  {volume} {95}},\ \bibinfo {pages} {075420} (\bibinfo {year}
  {2017})}\BibitemShut {NoStop}%
\bibitem [{\citenamefont {Tarnopolsky}\ \emph {et~al.}(2019)\citenamefont
  {Tarnopolsky}, \citenamefont {Kruchkov},\ and\ \citenamefont
  {Vishwanath}}]{Tarnopolsky_2019}%
  \BibitemOpen
  \bibfield  {author} {\bibinfo {author} {\bibfnamefont {G.}~\bibnamefont
  {Tarnopolsky}}, \bibinfo {author} {\bibfnamefont {A.~J.}\ \bibnamefont
  {Kruchkov}}, \ and\ \bibinfo {author} {\bibfnamefont {A.}~\bibnamefont
  {Vishwanath}},\ }\href {\doibase 10.1103/PhysRevLett.122.106405} {\bibfield
  {journal} {\bibinfo  {journal} {Phys. Rev. Lett.}\ }\textbf {\bibinfo
  {volume} {122}},\ \bibinfo {pages} {106405} (\bibinfo {year}
  {2019})}\BibitemShut {NoStop}%
\bibitem [{\citenamefont {Khalaf}\ \emph {et~al.}(2019)\citenamefont {Khalaf},
  \citenamefont {Kruchkov}, \citenamefont {Tarnopolsky},\ and\ \citenamefont
  {Vishwanath}}]{Khalaf_2019}%
  \BibitemOpen
  \bibfield  {author} {\bibinfo {author} {\bibfnamefont {E.}~\bibnamefont
  {Khalaf}}, \bibinfo {author} {\bibfnamefont {A.~J.}\ \bibnamefont
  {Kruchkov}}, \bibinfo {author} {\bibfnamefont {G.}~\bibnamefont
  {Tarnopolsky}}, \ and\ \bibinfo {author} {\bibfnamefont {A.}~\bibnamefont
  {Vishwanath}},\ }\href {\doibase 10.1103/PhysRevB.100.085109} {\bibfield
  {journal} {\bibinfo  {journal} {Phys. Rev. B}\ }\textbf {\bibinfo {volume}
  {100}},\ \bibinfo {pages} {085109} (\bibinfo {year} {2019})}\BibitemShut
  {NoStop}%
\bibitem [{\citenamefont {Li}\ \emph {et~al.}(2022)\citenamefont {Li},
  \citenamefont {Eaton}, \citenamefont {Fertig},\ and\ \citenamefont
  {Seradjeh}}]{Li2022}%
  \BibitemOpen
  \bibfield  {author} {\bibinfo {author} {\bibfnamefont {Y.}~\bibnamefont
  {Li}}, \bibinfo {author} {\bibfnamefont {A.}~\bibnamefont {Eaton}}, \bibinfo
  {author} {\bibfnamefont {H.~A.}\ \bibnamefont {Fertig}}, \ and\ \bibinfo
  {author} {\bibfnamefont {B.}~\bibnamefont {Seradjeh}},\ }\href {\doibase
  10.1103/PhysRevLett.128.026404} {\bibfield  {journal} {\bibinfo  {journal}
  {Phys. Rev. Lett.}\ }\textbf {\bibinfo {volume} {128}},\ \bibinfo {pages}
  {026404} (\bibinfo {year} {2022})}\BibitemShut {NoStop}%
\bibitem [{\citenamefont {Cao}\ \emph {et~al.}(2020)\citenamefont {Cao},
  \citenamefont {Chowdhury}, \citenamefont {Rodan-Legrain}, \citenamefont
  {Rubies-Bigorda}, \citenamefont {Watanabe}, \citenamefont {Taniguchi},
  \citenamefont {Senthil},\ and\ \citenamefont {Jarillo-Herrero}}]{Cao2020S}%
  \BibitemOpen
  \bibfield  {author} {\bibinfo {author} {\bibfnamefont {Y.}~\bibnamefont
  {Cao}}, \bibinfo {author} {\bibfnamefont {D.}~\bibnamefont {Chowdhury}},
  \bibinfo {author} {\bibfnamefont {D.}~\bibnamefont {Rodan-Legrain}}, \bibinfo
  {author} {\bibfnamefont {O.}~\bibnamefont {Rubies-Bigorda}}, \bibinfo
  {author} {\bibfnamefont {K.}~\bibnamefont {Watanabe}}, \bibinfo {author}
  {\bibfnamefont {T.}~\bibnamefont {Taniguchi}}, \bibinfo {author}
  {\bibfnamefont {T.}~\bibnamefont {Senthil}}, \ and\ \bibinfo {author}
  {\bibfnamefont {P.}~\bibnamefont {Jarillo-Herrero}},\ }\href {\doibase
  10.1103/PhysRevLett.124.076801} {\bibfield  {journal} {\bibinfo  {journal}
  {Phys. Rev. Lett.}\ }\textbf {\bibinfo {volume} {124}},\ \bibinfo {pages}
  {076801} (\bibinfo {year} {2020})}\BibitemShut {NoStop}%
\bibitem [{\citenamefont {Wu}\ \emph {et~al.}(2018{\natexlab{b}})\citenamefont
  {Wu}, \citenamefont {MacDonald},\ and\ \citenamefont {Martin}}]{Wu2018super}%
  \BibitemOpen
  \bibfield  {author} {\bibinfo {author} {\bibfnamefont {F.}~\bibnamefont
  {Wu}}, \bibinfo {author} {\bibfnamefont {A.~H.}\ \bibnamefont {MacDonald}}, \
  and\ \bibinfo {author} {\bibfnamefont {I.}~\bibnamefont {Martin}},\ }\href
  {\doibase 10.1103/PhysRevLett.121.257001} {\bibfield  {journal} {\bibinfo
  {journal} {Phys. Rev. Lett.}\ }\textbf {\bibinfo {volume} {121}},\ \bibinfo
  {pages} {257001} (\bibinfo {year} {2018}{\natexlab{b}})}\BibitemShut
  {NoStop}%
\bibitem [{\citenamefont {Lian}\ \emph {et~al.}(2019)\citenamefont {Lian},
  \citenamefont {Wang},\ and\ \citenamefont {Bernevig}}]{Lian2019}%
  \BibitemOpen
  \bibfield  {author} {\bibinfo {author} {\bibfnamefont {B.}~\bibnamefont
  {Lian}}, \bibinfo {author} {\bibfnamefont {Z.}~\bibnamefont {Wang}}, \ and\
  \bibinfo {author} {\bibfnamefont {B.~A.}\ \bibnamefont {Bernevig}},\ }\href
  {\doibase 10.1103/PhysRevLett.122.257002} {\bibfield  {journal} {\bibinfo
  {journal} {Phys. Rev. Lett.}\ }\textbf {\bibinfo {volume} {122}},\ \bibinfo
  {pages} {257002} (\bibinfo {year} {2019})}\BibitemShut {NoStop}%
\bibitem [{\citenamefont {Gonz\'alez}\ and\ \citenamefont
  {Stauber}(2019)}]{Gonzalez2019}%
  \BibitemOpen
  \bibfield  {author} {\bibinfo {author} {\bibfnamefont {J.}~\bibnamefont
  {Gonz\'alez}}\ and\ \bibinfo {author} {\bibfnamefont {T.}~\bibnamefont
  {Stauber}},\ }\href {\doibase 10.1103/PhysRevLett.122.026801} {\bibfield
  {journal} {\bibinfo  {journal} {Phys. Rev. Lett.}\ }\textbf {\bibinfo
  {volume} {122}},\ \bibinfo {pages} {026801} (\bibinfo {year}
  {2019})}\BibitemShut {NoStop}%
\bibitem [{\citenamefont {Christos}\ \emph {et~al.}(2022)\citenamefont
  {Christos}, \citenamefont {Sachdev},\ and\ \citenamefont
  {Scheurer}}]{Christos2022}%
  \BibitemOpen
  \bibfield  {author} {\bibinfo {author} {\bibfnamefont {M.}~\bibnamefont
  {Christos}}, \bibinfo {author} {\bibfnamefont {S.}~\bibnamefont {Sachdev}}, \
  and\ \bibinfo {author} {\bibfnamefont {M.~S.}\ \bibnamefont {Scheurer}},\
  }\href {\doibase 10.1103/PhysRevX.12.021018} {\bibfield  {journal} {\bibinfo
  {journal} {Phys. Rev. X}\ }\textbf {\bibinfo {volume} {12}},\ \bibinfo
  {pages} {021018} (\bibinfo {year} {2022})}\BibitemShut {NoStop}%
\bibitem [{\citenamefont {Fischer}\ \emph {et~al.}(2022)\citenamefont
  {Fischer}, \citenamefont {Goodwin}, \citenamefont {Mostofi}, \citenamefont
  {Lischner}, \citenamefont {Kennes},\ and\ \citenamefont
  {Klebl}}]{Fischer2022}%
  \BibitemOpen
  \bibfield  {author} {\bibinfo {author} {\bibfnamefont {A.}~\bibnamefont
  {Fischer}}, \bibinfo {author} {\bibfnamefont {Z.~A.~H.}\ \bibnamefont
  {Goodwin}}, \bibinfo {author} {\bibfnamefont {A.~A.}\ \bibnamefont
  {Mostofi}}, \bibinfo {author} {\bibfnamefont {J.}~\bibnamefont {Lischner}},
  \bibinfo {author} {\bibfnamefont {D.~M.}\ \bibnamefont {Kennes}}, \ and\
  \bibinfo {author} {\bibfnamefont {L.}~\bibnamefont {Klebl}},\ }\href
  {\doibase 10.1038/s41535-021-00410-w} {\bibfield  {journal} {\bibinfo
  {journal} {npj Quantum Materials}\ }\textbf {\bibinfo {volume} {7}},\
  \bibinfo {pages} {5} (\bibinfo {year} {2022})}\BibitemShut {NoStop}%
\bibitem [{\citenamefont {Song}\ and\ \citenamefont
  {Bernevig}(2022)}]{Song2022}%
  \BibitemOpen
  \bibfield  {author} {\bibinfo {author} {\bibfnamefont {Z.-D.}\ \bibnamefont
  {Song}}\ and\ \bibinfo {author} {\bibfnamefont {B.~A.}\ \bibnamefont
  {Bernevig}},\ }\href {\doibase 10.1103/PhysRevLett.129.047601} {\bibfield
  {journal} {\bibinfo  {journal} {Phys. Rev. Lett.}\ }\textbf {\bibinfo
  {volume} {129}},\ \bibinfo {pages} {047601} (\bibinfo {year}
  {2022})}\BibitemShut {NoStop}%
\bibitem [{\citenamefont {Tilak}\ \emph {et~al.}(2021)\citenamefont {Tilak},
  \citenamefont {Lai}, \citenamefont {Wu}, \citenamefont {Zhang}, \citenamefont
  {Xu}, \citenamefont {Ribeiro}, \citenamefont {Canfield},\ and\ \citenamefont
  {Andrei}}]{Tilak2021}%
  \BibitemOpen
  \bibfield  {author} {\bibinfo {author} {\bibfnamefont {N.}~\bibnamefont
  {Tilak}}, \bibinfo {author} {\bibfnamefont {X.}~\bibnamefont {Lai}}, \bibinfo
  {author} {\bibfnamefont {S.}~\bibnamefont {Wu}}, \bibinfo {author}
  {\bibfnamefont {Z.}~\bibnamefont {Zhang}}, \bibinfo {author} {\bibfnamefont
  {M.}~\bibnamefont {Xu}}, \bibinfo {author} {\bibfnamefont {R.~d.~A.}\
  \bibnamefont {Ribeiro}}, \bibinfo {author} {\bibfnamefont {P.~C.}\
  \bibnamefont {Canfield}}, \ and\ \bibinfo {author} {\bibfnamefont {E.~Y.}\
  \bibnamefont {Andrei}},\ }\href {\doibase 10.1038/s41467-021-24480-3}
  {\bibfield  {journal} {\bibinfo  {journal} {Nature Communications}\ }\textbf
  {\bibinfo {volume} {12}},\ \bibinfo {pages} {4180} (\bibinfo {year}
  {2021})}\BibitemShut {NoStop}%
\bibitem [{\citenamefont {Chou}\ and\ \citenamefont
  {Sarma}(2022)}]{chou2022kondo}%
  \BibitemOpen
  \bibfield  {author} {\bibinfo {author} {\bibfnamefont {Y.-Z.}\ \bibnamefont
  {Chou}}\ and\ \bibinfo {author} {\bibfnamefont {S.~D.}\ \bibnamefont
  {Sarma}},\ }\href@noop {} {\enquote {\bibinfo {title} {Kondo lattice model in
  magic-angle twisted bilayer graphene},}\ } (\bibinfo {year} {2022}),\ \Eprint
  {http://arxiv.org/abs/2211.15682} {arXiv:2211.15682 [cond-mat.str-el]}
  \BibitemShut {NoStop}%
\bibitem [{\citenamefont {Yu}\ \emph {et~al.}(2023)\citenamefont {Yu},
  \citenamefont {Xie}, \citenamefont {Bernevig},\ and\ \citenamefont
  {Sarma}}]{yu2023magicangle}%
  \BibitemOpen
  \bibfield  {author} {\bibinfo {author} {\bibfnamefont {J.}~\bibnamefont
  {Yu}}, \bibinfo {author} {\bibfnamefont {M.}~\bibnamefont {Xie}}, \bibinfo
  {author} {\bibfnamefont {B.~A.}\ \bibnamefont {Bernevig}}, \ and\ \bibinfo
  {author} {\bibfnamefont {S.~D.}\ \bibnamefont {Sarma}},\ }\href@noop {}
  {\enquote {\bibinfo {title} {Magic-angle twisted symmetric trilayer graphene
  as topological heavy fermion problem},}\ } (\bibinfo {year} {2023}),\ \Eprint
  {http://arxiv.org/abs/2301.04171} {arXiv:2301.04171 [cond-mat.mes-hall]}
  \BibitemShut {NoStop}%
\bibitem [{\citenamefont {Hu}\ \emph {et~al.}(2023{\natexlab{a}})\citenamefont
  {Hu}, \citenamefont {Rai}, \citenamefont {Crippa}, \citenamefont
  {Herzog-Arbeitman}, \citenamefont {C?lug?ru}, \citenamefont {Wehling},
  \citenamefont {Sangiovanni}, \citenamefont {Valenti}, \citenamefont
  {Tsvelik},\ and\ \citenamefont {Bernevig}}]{hu2023symmetric}%
  \BibitemOpen
  \bibfield  {author} {\bibinfo {author} {\bibfnamefont {H.}~\bibnamefont
  {Hu}}, \bibinfo {author} {\bibfnamefont {G.}~\bibnamefont {Rai}}, \bibinfo
  {author} {\bibfnamefont {L.}~\bibnamefont {Crippa}}, \bibinfo {author}
  {\bibfnamefont {J.}~\bibnamefont {Herzog-Arbeitman}}, \bibinfo {author}
  {\bibfnamefont {D.}~\bibnamefont {C?lug?ru}}, \bibinfo {author}
  {\bibfnamefont {T.}~\bibnamefont {Wehling}}, \bibinfo {author} {\bibfnamefont
  {G.}~\bibnamefont {Sangiovanni}}, \bibinfo {author} {\bibfnamefont
  {R.}~\bibnamefont {Valenti}}, \bibinfo {author} {\bibfnamefont {A.~M.}\
  \bibnamefont {Tsvelik}}, \ and\ \bibinfo {author} {\bibfnamefont {B.~A.}\
  \bibnamefont {Bernevig}},\ }\href@noop {} {\enquote {\bibinfo {title}
  {Symmetric kondo lattice states in doped strained twisted bilayer
  graphene},}\ } (\bibinfo {year} {2023}{\natexlab{a}}),\ \Eprint
  {http://arxiv.org/abs/2301.04673} {arXiv:2301.04673 [cond-mat.str-el]}
  \BibitemShut {NoStop}%
\bibitem [{\citenamefont {Hu}\ \emph {et~al.}(2023{\natexlab{b}})\citenamefont
  {Hu}, \citenamefont {Bernevig},\ and\ \citenamefont {Tsvelik}}]{hu2023kondo}%
  \BibitemOpen
  \bibfield  {author} {\bibinfo {author} {\bibfnamefont {H.}~\bibnamefont
  {Hu}}, \bibinfo {author} {\bibfnamefont {B.~A.}\ \bibnamefont {Bernevig}}, \
  and\ \bibinfo {author} {\bibfnamefont {A.~M.}\ \bibnamefont {Tsvelik}},\
  }\href@noop {} {\enquote {\bibinfo {title} {Kondo lattice model of
  magic-angle twisted-bilayer graphene: Hund's rule, local-moment fluctuations,
  and low-energy effective theory},}\ } (\bibinfo {year}
  {2023}{\natexlab{b}}),\ \Eprint {http://arxiv.org/abs/2301.04669}
  {arXiv:2301.04669 [cond-mat.str-el]} \BibitemShut {NoStop}%
\bibitem [{\citenamefont {Zhou}\ and\ \citenamefont
  {Song}(2023)}]{zhou2023kondo}%
  \BibitemOpen
  \bibfield  {author} {\bibinfo {author} {\bibfnamefont {G.-D.}\ \bibnamefont
  {Zhou}}\ and\ \bibinfo {author} {\bibfnamefont {Z.-D.}\ \bibnamefont
  {Song}},\ }\href@noop {} {\enquote {\bibinfo {title} {Kondo phase in twisted
  bilayer graphene -- a unified theory for distinct experiments},}\ } (\bibinfo
  {year} {2023}),\ \Eprint {http://arxiv.org/abs/2301.04661} {arXiv:2301.04661
  [cond-mat.str-el]} \BibitemShut {NoStop}%
\bibitem [{\citenamefont {Huang}\ \emph {et~al.}(2023)\citenamefont {Huang},
  \citenamefont {Zhang}, \citenamefont {Pan}, \citenamefont {Li}, \citenamefont
  {Sun}, \citenamefont {Dai},\ and\ \citenamefont {Meng}}]{huang2023evolution}%
  \BibitemOpen
  \bibfield  {author} {\bibinfo {author} {\bibfnamefont {C.}~\bibnamefont
  {Huang}}, \bibinfo {author} {\bibfnamefont {X.}~\bibnamefont {Zhang}},
  \bibinfo {author} {\bibfnamefont {G.}~\bibnamefont {Pan}}, \bibinfo {author}
  {\bibfnamefont {H.}~\bibnamefont {Li}}, \bibinfo {author} {\bibfnamefont
  {K.}~\bibnamefont {Sun}}, \bibinfo {author} {\bibfnamefont {X.}~\bibnamefont
  {Dai}}, \ and\ \bibinfo {author} {\bibfnamefont {Z.}~\bibnamefont {Meng}},\
  }\href@noop {} {\enquote {\bibinfo {title} {Evolution from quantum anomalous
  hall insulator to heavy-fermion semimetal in magic-angle twisted bilayer
  graphene},}\ } (\bibinfo {year} {2023}),\ \Eprint
  {http://arxiv.org/abs/2304.14064} {arXiv:2304.14064 [cond-mat.str-el]}
  \BibitemShut {NoStop}%
\bibitem [{\citenamefont {C?lug?ru}\ \emph {et~al.}(2023)\citenamefont
  {C?lug?ru}, \citenamefont {Borovkov}, \citenamefont {Lau}, \citenamefont
  {Coleman}, \citenamefont {Song},\ and\ \citenamefont
  {Bernevig}}]{calugaru2023tbg}%
  \BibitemOpen
  \bibfield  {author} {\bibinfo {author} {\bibfnamefont {D.}~\bibnamefont
  {C?lug?ru}}, \bibinfo {author} {\bibfnamefont {M.}~\bibnamefont {Borovkov}},
  \bibinfo {author} {\bibfnamefont {L.~L.~H.}\ \bibnamefont {Lau}}, \bibinfo
  {author} {\bibfnamefont {P.}~\bibnamefont {Coleman}}, \bibinfo {author}
  {\bibfnamefont {Z.-D.}\ \bibnamefont {Song}}, \ and\ \bibinfo {author}
  {\bibfnamefont {B.~A.}\ \bibnamefont {Bernevig}},\ }\href@noop {} {\enquote
  {\bibinfo {title} {Tbg as topological heavy fermion: Ii. analytical
  approximations of the model parameters},}\ } (\bibinfo {year} {2023}),\
  \Eprint {http://arxiv.org/abs/2303.03429} {arXiv:2303.03429
  [cond-mat.str-el]} \BibitemShut {NoStop}%
\bibitem [{\citenamefont {Singh}\ \emph {et~al.}(2023)\citenamefont {Singh},
  \citenamefont {Chew}, \citenamefont {Herzog-Arbeitman}, \citenamefont
  {Bernevig},\ and\ \citenamefont {Vafek}}]{singh2023topological}%
  \BibitemOpen
  \bibfield  {author} {\bibinfo {author} {\bibfnamefont {K.}~\bibnamefont
  {Singh}}, \bibinfo {author} {\bibfnamefont {A.}~\bibnamefont {Chew}},
  \bibinfo {author} {\bibfnamefont {J.}~\bibnamefont {Herzog-Arbeitman}},
  \bibinfo {author} {\bibfnamefont {B.~A.}\ \bibnamefont {Bernevig}}, \ and\
  \bibinfo {author} {\bibfnamefont {O.}~\bibnamefont {Vafek}},\ }\href@noop {}
  {\enquote {\bibinfo {title} {Topological heavy fermions in magnetic field},}\
  } (\bibinfo {year} {2023}),\ \Eprint {http://arxiv.org/abs/2305.08171}
  {arXiv:2305.08171 [cond-mat.str-el]} \BibitemShut {NoStop}%
\bibitem [{\citenamefont {Chou}\ and\ \citenamefont
  {Sarma}(2023)}]{chou2023scaling}%
  \BibitemOpen
  \bibfield  {author} {\bibinfo {author} {\bibfnamefont {Y.-Z.}\ \bibnamefont
  {Chou}}\ and\ \bibinfo {author} {\bibfnamefont {S.~D.}\ \bibnamefont
  {Sarma}},\ }\href@noop {} {\enquote {\bibinfo {title} {Scaling theory of
  intrinsic kondo and hund's rule interactions in magic-angle twisted bilayer
  graphene},}\ } (\bibinfo {year} {2023}),\ \Eprint
  {http://arxiv.org/abs/2306.03121} {arXiv:2306.03121 [cond-mat.str-el]}
  \BibitemShut {NoStop}%
\bibitem [{\citenamefont {Oh}\ \emph {et~al.}(2021)\citenamefont {Oh},
  \citenamefont {Nuckolls}, \citenamefont {Wong}, \citenamefont {Lee},
  \citenamefont {Liu}, \citenamefont {Watanabe}, \citenamefont {Taniguchi},\
  and\ \citenamefont {Yazdani}}]{Oh2021}%
  \BibitemOpen
  \bibfield  {author} {\bibinfo {author} {\bibfnamefont {M.}~\bibnamefont
  {Oh}}, \bibinfo {author} {\bibfnamefont {K.~P.}\ \bibnamefont {Nuckolls}},
  \bibinfo {author} {\bibfnamefont {D.}~\bibnamefont {Wong}}, \bibinfo {author}
  {\bibfnamefont {R.~L.}\ \bibnamefont {Lee}}, \bibinfo {author} {\bibfnamefont
  {X.}~\bibnamefont {Liu}}, \bibinfo {author} {\bibfnamefont {K.}~\bibnamefont
  {Watanabe}}, \bibinfo {author} {\bibfnamefont {T.}~\bibnamefont {Taniguchi}},
  \ and\ \bibinfo {author} {\bibfnamefont {A.}~\bibnamefont {Yazdani}},\ }\href
  {\doibase 10.1038/s41586-021-04121-x} {\bibfield  {journal} {\bibinfo
  {journal} {Nature}\ }\textbf {\bibinfo {volume} {600}},\ \bibinfo {pages}
  {240} (\bibinfo {year} {2021})}\BibitemShut {NoStop}%
\bibitem [{\citenamefont {Lavagna}\ \emph {et~al.}(1987)\citenamefont
  {Lavagna}, \citenamefont {Millis},\ and\ \citenamefont {Lee}}]{Lavagna1987}%
  \BibitemOpen
  \bibfield  {author} {\bibinfo {author} {\bibfnamefont {M.}~\bibnamefont
  {Lavagna}}, \bibinfo {author} {\bibfnamefont {A.~J.}\ \bibnamefont {Millis}},
  \ and\ \bibinfo {author} {\bibfnamefont {P.~A.}\ \bibnamefont {Lee}},\ }\href
  {\doibase 10.1103/PhysRevLett.58.266} {\bibfield  {journal} {\bibinfo
  {journal} {Phys. Rev. Lett.}\ }\textbf {\bibinfo {volume} {58}},\ \bibinfo
  {pages} {266} (\bibinfo {year} {1987})}\BibitemShut {NoStop}%
\bibitem [{\citenamefont {Dorin}\ and\ \citenamefont
  {Schlottmann}(1993)}]{Dorin1993}%
  \BibitemOpen
  \bibfield  {author} {\bibinfo {author} {\bibfnamefont {V.}~\bibnamefont
  {Dorin}}\ and\ \bibinfo {author} {\bibfnamefont {P.}~\bibnamefont
  {Schlottmann}},\ }\href {\doibase 10.1103/PhysRevB.47.5095} {\bibfield
  {journal} {\bibinfo  {journal} {Phys. Rev. B}\ }\textbf {\bibinfo {volume}
  {47}},\ \bibinfo {pages} {5095} (\bibinfo {year} {1993})}\BibitemShut
  {NoStop}%
\bibitem [{\citenamefont {Lau}\ and\ \citenamefont
  {Coleman}(2023)}]{lau2023topological}%
  \BibitemOpen
  \bibfield  {author} {\bibinfo {author} {\bibfnamefont {L.~L.~H.}\
  \bibnamefont {Lau}}\ and\ \bibinfo {author} {\bibfnamefont {P.}~\bibnamefont
  {Coleman}},\ }\href@noop {} {\enquote {\bibinfo {title} {Topological mixed
  valence model for twisted bilayer graphene},}\ } (\bibinfo {year} {2023}),\
  \Eprint {http://arxiv.org/abs/2303.02670} {arXiv:2303.02670
  [cond-mat.str-el]} \BibitemShut {NoStop}%
\end{thebibliography}%
%

\vspace{-5mm}
\onecolumngrid
\newpage

\renewcommand{\thefigure}{S\arabic{figure}}
\renewcommand{\theequation}{S\arabic{equation}}
\setcounter{equation}{0}
\setcounter{figure}{0}

\title{
Supplemental Material for ``Topological Mixed Valence Model in Magic-Angle Twisted Bilayer Graphene''
}

\begin{abstract}
\end{abstract}

{
\let\clearpage\relax
\maketitle
}


\onecolumngrid
\section{Hamiltonian and Parameters}
The periodic Anderson model in twisted bilayer graphene \cite{Song2022} can be expressed as 
\begin{equation}
	\hat{H}=\hat{H}_{0,c}+\hat{H}_{0,f}+\hat{H}_{0,cf}+\hat{H}_{U},
\end{equation}
where $\hat{H}_{0,c}$ is the Hamiltonian of the conduction ($c$-) electrons in AB/BA moir\'e lattice sites, $\hat{H}_{0,f}$ is the Hamiltonian of flat band ($f$-) electrons in AA moir\'e lattice sites, $\hat{H}_{0,cf}$ is the coupling Hamiltonian of them, and $\hat{H}_{U}$ is the Hamiltonian of Coulomb interaction among the $f$- electrons. The explicit form of them are
\begin{equation}
	\hat{H}_{0,c}=\sum_{a,a^{\prime},\eta,s}\sum_{\bold{p}} h^{(c,\eta s)}_{a a^{\prime}}(\bold{p}) c^{\dagger}_{\bold{p},a,\eta,s} c_{\bold{p},a^{\prime},\eta,s}=\sum_{a,a^{\prime},\eta,s}\sum_{\bold{G}}\sum_{\bold{k}\in \text{mBZ}}  h^{(c,\eta s)}_{a a^{\prime}}(\bold{k}+\bold{G}) c^{\dagger}_{\bold{k}+\bold{G},a,\eta,s} c_{\bold{k}+\bold{G},a^{\prime},\eta,s},
\end{equation}

\begin{equation}
	\hat{H}_{0,f}=\sum_{\alpha,\alpha^{\prime},\eta,s}\sum_{\bold{k}} h^{(f,\eta s)}_{\alpha \alpha^{\prime}}(\bold{k}) f^{\dagger}_{\bold{k},\alpha,\eta,s} f_{\bold{k},\alpha^{\prime},\eta,s},
\end{equation}

\begin{equation}
	\hat{H}_{0,cf}=\sum_{\alpha,a,\eta,s}\sum_{\bold{G}}\sum_{\bold{k}\in \text{mBZ}} [V^{(\eta s)}_{\alpha a}(\bold{k}+\bold{G}) f^{\dagger}_{\bold{k},\alpha,\eta,s} c_{\bold{k}+\bold{G},a,\eta,s}+\text{H.c.}],
\end{equation}

and
\begin{equation}
	\hat{H}_{U}=\frac{U}{2}\sum_{\bold{R}}:\hat{n}^{f}_{\bold{R}}::\hat{n}^{f}_{\bold{R}}:,
\end{equation}
where $\bold{p}=\bold{k}+\bold{G}$, $\hat{n}^{f}_{\bold{R}}$ is the on-site density operator of $f$-electrons and $U$ is Coulomb interaction among them.
The single particle Hamiltonian of the model can also be expressed as 
\begin{equation}
\begin{aligned}
\hat{H_0}&=
\begin{bmatrix}
h^{(c,\eta s)}(\bold{p}) & V^{(\eta s)}(\bold{p}) \\
(V^{(\eta s)}(\bold{p}))^{\dagger}  & h^{(f,\eta s)}(\bold{k})\\
\end{bmatrix}\\
&=
\begin{bmatrix}
0_{2\times2}-\mu \sigma_0 & \nu_\star(\eta p_x\sigma_0+ip_y\sigma_z) & \gamma\sigma_0+\nu_\star^{\prime}(\eta p_x\sigma_x+p_y\sigma_y)\\
\nu_\star(\eta p_x\sigma_0-ip_y\sigma_z)  & M\sigma_x-\mu\sigma_0 & 0_{2\times2}\\
\gamma\sigma_0+\nu_\star^{\prime}(\eta p_x\sigma_x+p_y\sigma_y) & 0_{2\times2} & (\epsilon_{f_0}-\mu)\sigma_0
\end{bmatrix}\\
\\
\end{aligned},
\end{equation}
where
\begin{equation}
\begin{aligned}
h^{(c,\eta s)}(\bold{p})&=
\begin{bmatrix}
0_{2\times2}-\mu \sigma_0 & \nu_\star(\eta p_x\sigma_0+ip_y\sigma_z) \\
\nu_\star(\eta p_x\sigma_0-ip_y\sigma_z)  & M\sigma_x-\mu\sigma_0 \\
\end{bmatrix},
\end{aligned}
\end{equation}

\begin{equation}
\begin{aligned}
V^{(\eta s)}(\bold{p})&=e^{\frac{-|\bold{p}|^{2}\lambda^{2}}{2}}
\begin{bmatrix}
\gamma\sigma_0+\nu_\star^{\prime}(\eta p_x\sigma_x+p_y\sigma_y)\\
0_{2\times2}\\
\end{bmatrix},
\end{aligned}
\end{equation}
and $h^{(f,\eta s)}=(\epsilon_{f_0}-\mu)\sigma_0$.

\begin{figure}[t]
   \centering
   \includegraphics[width=3.in]{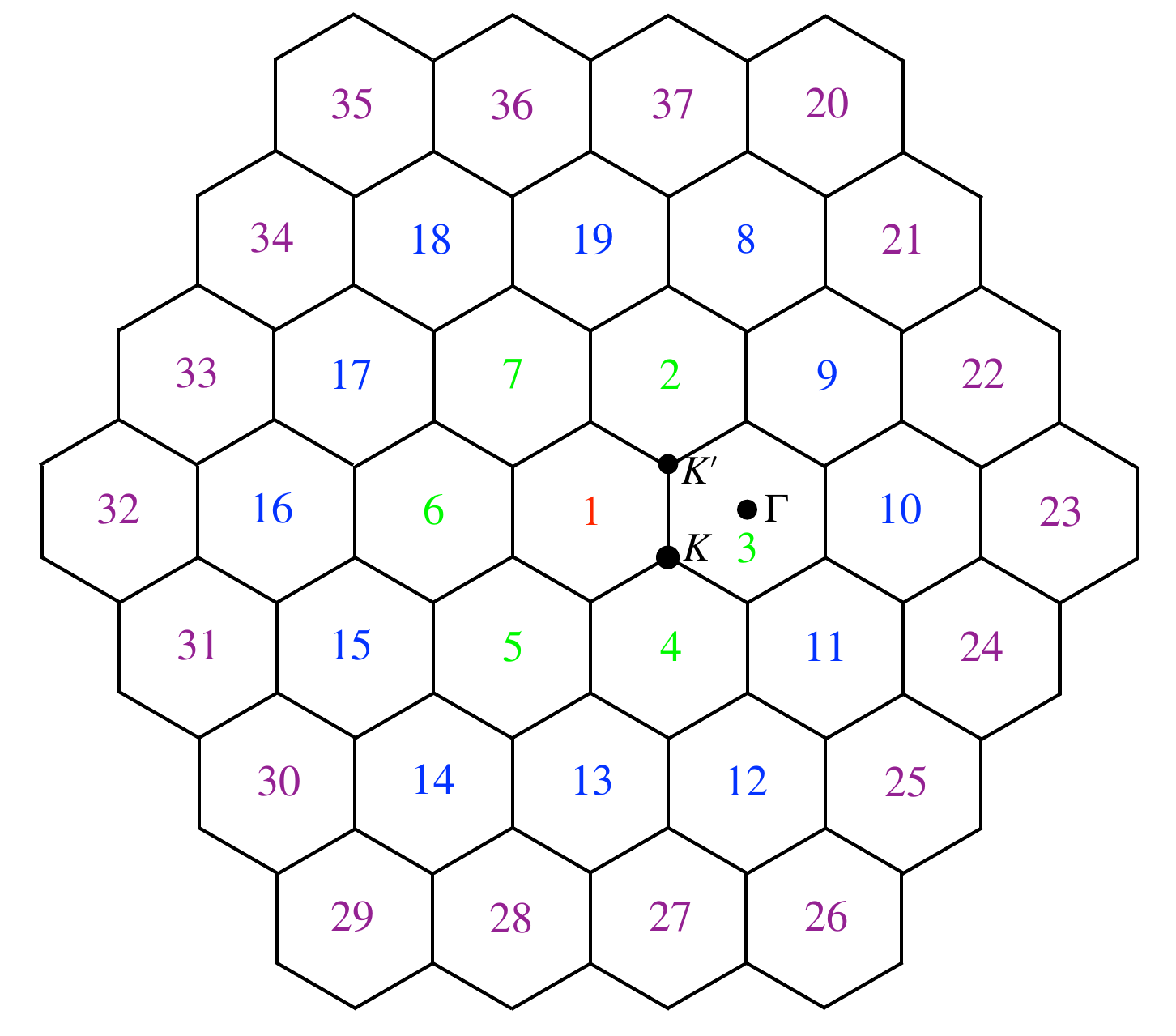} 
   \caption{Sketch of moir\'e momentum space which includes $N_G=37$ moir\'e Brillouin zones. $K$, $K^{\prime}$ and $\Gamma$ are the symmetry points in the moir\'e Brillouin zone. In numerical calculation, we increase $N_G$ from $9$ to $19$ and $37$ to check the convergence.}
   \label{FigS1:sketch}
\end{figure}

The Hamiltonian is expressed in the moir\'e momentum space using the plane wave approximation. The size of moir\'e momentum space (per valley per spin) is $2+4N_G$ which means $2$ flat bands and $4$ conduction bands and conduction bands are expanded to $N_G$ moir\'e Brillouin zone (mBZ), see Fig.~\ref{FigS1:sketch}. Then the single particle Hamiltonian can be written in matrix form as
\begin{equation}
\begin{aligned}
\hat{H}_{0}= \left(\begin{array}{cccc}\hat{H}_{0,c} & \hat{H}_{0,cf} \\
(\hat{H}_{0,cf})^{\dagger} & \hat{H}_{0,f}  \\ 
  \end{array}\right),
\end{aligned}
\end{equation}
where
\begin{equation}
\begin{aligned}
\hat{H}_{0,c}= \left(\begin{array}{cccc}h^{(c,\eta s)}(\bold{k}+\bold{G_1}) & 0 & \cdots & 0 \\
0 & h^{(c,\eta s)}(\bold{k}+\bold{G_2}) & \cdots & 0 \\ 
 \vdots & \vdots & \ddots & \vdots \\
 0 & 0 & \cdots & h^{(c,\eta s)}(\bold{k}+\bold{G_{N_{G}}}) \ \end{array}\right),
\end{aligned}
\end{equation}

\begin{equation}
\begin{aligned}
\hat{H}_{0,cf}= \left(\begin{array}{cccc}V^{(\eta s)}_{\alpha a}(\bold{k}+\bold{G_1}) & V^{(\eta s)}_{\alpha a}(\bold{k}+\bold{G_2}) & \cdots & V^{(\eta s)}_{\alpha a}(\bold{k}+\bold{G_{N_G}})  \ \end{array}\right),
\end{aligned}
\end{equation}

and $\hat{H}_{0,f}=h^{(f,\eta s)}.$ Note that $\bold{G_{n}=\Gamma_{n}}$ and $\bold{\Gamma_{n}}$ is the position of symmetry point in $n$-th moir\'e Brillouin zone, for example, $\bold{G_{1}=\Gamma_{1}}=0$.

The values of the parameters are $\nu_\star=-4.303$ eV\AA, $M=3.697$ meV, $\gamma=-24.75$ meV, $\nu_\star^{\prime}=1.622$ eV\AA, and $\lambda=0.3375$ $a_{M}$, where $a_{M}$ is the moir\'e lattice constant. Note that all these parameters are corresponding to $\theta_m=1.05^{\circ}$, $U_0=W_{AA}/W_{AB}=0.8$, and the velocity of electron in single layer graphene $v_F=5.94$ eV\AA.

\section{Slave-boson Mean field equations in Large-N Expansion}

We extend the number of orbital, spin, and valley for both $c$- and $f$-electrons to $N$ and set
\begin{equation}
	U=\infty .
\end{equation}
By introducing the slave-boson operators $b^{\dagger}_{\bold{R}}$ and $b_{\bold{R}}$ at each AA site to exclude the double occupancy. The Hamiltonian in the large-N approximation is 
\begin{equation}
\begin{aligned}
	\hat{H}&=\hat{H}_{0,c}+\hat{H}_{0,f}+\hat{H}_{0, cf}\\
	&=\sum_{l}^{N}\left\{\sum_{a,a^{\prime}}\sum_{\bold{p}} h^{(c)}_{a a^{\prime}}(\bold{p}) c^{\dagger}_{\bold{p},a, l} c_{\bold{p},a^{\prime},l}+\sum_{\alpha,\alpha^{\prime}}\sum_{\bold{k}} h^{(f)}_{\alpha \alpha^{\prime}}(\bold{k}) f^{\dagger}_{\bold{k},\alpha, l} f_{\bold{k},\alpha^{\prime}, l}\right\}\\
	&+\frac{1}{\sqrt{N_{L}}}\sum_{l}^{N}\left\{\sum_{\alpha,a}\sum_{\bold{G}}\sum_{\bold{k}\in \text{mBZ}}\sum_{\bold{q}} [V_{\alpha a}(\bold{k}+\bold{G}) c^{\dagger}_{\bold{k}+\bold{G},a, l}f_{\bold{k}+\bold{q},\alpha, l}b^{\dagger}_{\bold{q}}+\text{H.c.}]\right\}. 
\end{aligned}
\end{equation}
The constraint is 
\begin{equation}
	Q=\sum^{N}_{l=1}\sum_{\alpha}(f^{\dagger}_{\bold{R},\alpha, l} f_{\bold{R},\alpha, l}+b^{\dagger}_{\bold{R}}b_{\bold{R}}).
\end{equation}
Note that $N$ includes orbital, spin, and valley. Here, we suppose different valley has the same band structure and we choose the valley $\eta=+$ in the above Hamiltonian. In our case, $N=4$ and $N=\infty$ is the mean field value. We stress that there are two index spaces in the Hamiltonian. One is the extended $N$ space with $1/N$ expansion corresponding to the index $l$. Another is the moir\'e momentum space which has size of $N_G$ corresponding to the index $a, a^{\prime}, \alpha$, and $\alpha^{\prime}$. We only consider one species of the slave boson for simplification. Since we have two flat bands, two species of the slave bosons \cite{Dorin1993} might be needed and we will consider this case in future works.

We introduce Lagrangian multipliers $\lambda_{\bold{R}}$ to ensure the number of $f$ electrons is $Q$ or $N_f$, where $Q\equiv N_f$. We rewrite $Q\rightarrow q_0 N$, $b_{\bold{R}}\rightarrow b_{\bold{R}} \sqrt{N}$, and $V_{\alpha a}\rightarrow V_{\alpha a}/ \sqrt{N}$. The local gauge transformation is $b_{\bold{R}}=\rho_{\bold{R}}\exp(i\theta_{\bold{R}})$, $f_{\bold{R}}=f^{\prime}_{\bold{R}}\exp(i\theta_{\bold{R}})$ and $\lambda_{\bold{R}}=\lambda^{\prime}_{\bold{R}}-\theta_{\bold{R}}$. We then rewrite $f^{\prime}_{\bold{R}}$ and $\lambda^{\prime}_{\bold{R}}$ to $f_{\bold{R}}$ and $\lambda_{\bold{R}}$. The partition function is 
\begin{equation}
      Z=\int \mathscr{D}(cc^{\dagger}ff^{\dagger}\rho\lambda)\exp(-S),	
\end{equation}
where the action is $S=\int_0^{\beta} L d\tau$
\begin{equation}
\begin{aligned}
	& L
	=\sum^{N}_{\substack{a,a^{\prime};l=1\\
                  \bold{p}}}(\partial_{\tau}+ h^{(c)}_{a a^{\prime}}(\bold{p})) c^{\dagger}_{\bold{p},a, l}c_{\bold{p},a^{\prime},l}(\tau)
	+\sum_{\substack{\alpha,\alpha^{\prime}\\
                  \bold{\bold{k},\bold{k}^{\prime}\in \text{mBZ}}}}(\partial_{\tau}+h^{(f)}_{\alpha \alpha^{\prime}}(\bold{k})\delta_{\bold{k},\bold{k}^{\prime}}+\frac{i\lambda(\bold{k}-\bold{k}^{\prime};\tau)}{\sqrt{N_{L}}}) f^{\dagger}_{\bold{k},\alpha, l} f_{\bold{k}^{\prime},\alpha^{\prime}, l}(\tau)+\frac{1}{\sqrt{N_{L}}}\sum^{N}_{\substack{\bold{G},\alpha,a;l=1\\
                  \bold{\bold{k},\bold{k}^{\prime}\in \text{mBZ}}}}
	\\&\left.[V_{\alpha a}(\bold{k}+\bold{G}) c^{\dagger}_{\bold{k}+\bold{G},a, l}(\tau)f_{\bold{k}^{\prime},\alpha, l}(\tau)\rho^{\dagger}(\bold{k}-\bold{k}^{\prime};\tau)+\text{H.c.}]\right.
	+\left.\frac{iN}{\sqrt{N_{L}}}\sum_{\bold{k},\bold{k}^{\prime}\in \text{mBZ}}  \rho({\bold{k}};\tau)\lambda(\bold{k}^{\prime}-\bold{k};\tau)\rho(-\bold{k}^{\prime};\tau)\right.\left.-iq_0N\sqrt{N_L}\lambda(0;\tau)\right.\\, 
\end{aligned}
\end{equation}       
where $\beta=\frac{1}{T}$ and $T$ is the temperature.
We set $\lambda(\bold{k;\tau})=\frac{1}{T}\bar{\lambda}\delta_{\bold{k},0}$, $\rho(\bold{k};\tau)=\frac{1}{T}\bar{\rho}\delta_{\bold{k},0}$, then the mean field action is
\begin{equation}
    \begin{aligned}
    S_0=&-N\sum_{\bold{k}\in \text{mBZ}}\sum_{\omega_n} \ln^{\det{(-i\omega_n+A_0)}}+\frac{iN}{T}(\bar{\rho}^2/\sqrt{N_L}-q_0\sqrt{N_L})\bar{\lambda}\\
    =&-N\sum_{\bold{k}\in \text{mBZ}}\sum_{\omega_n}\sum_{j=1}^{2+4N_G} \ln^{(-i\omega_n+\mathcal{E}_j)}+\frac{iN}{T}(\bar{\rho}^2/\sqrt{N_L}-q_0\sqrt{N_L})\bar{\lambda}\\
    =&-N\sum_{\bold{k}\in \text{mBZ}}\sum_{j=1}^{2+4N_G} \ln^{(1+e^{\mathcal{E}_j/T})}+\frac{iN}{T}(\bar{\rho}^2/\sqrt{N_L}-q_0\sqrt{N_L})\bar{\lambda},
    \end{aligned}	
\end{equation}
where $\mathcal{E}_j$ is the eigenvalues of $A_0$ and $A_0$ is a matrix with size of $2+4N_G$. We rewrite $\frac{\bar{\lambda}}{\sqrt{N_{L}}}\rightarrow \bar{\lambda}$ and $\frac{\bar{\rho}}{\sqrt{N_{L}}}\rightarrow \bar{\rho}$, then we have
\begin{equation}
    \begin{aligned}
    S_0=&-N\sum_{\bold{k}\in \text{mBZ}}\sum_{\omega_n} \ln^{\det{(-i\omega_n+A_0)}}+\frac{iN}{T}N_L(\bar{\rho}^2-q_0)\bar{\lambda}\\
    =&-N\sum_{\bold{k}\in \text{mBZ}}\sum_{\omega_n}\sum_{j=1}^{2+4N_G} \ln^{(-i\omega_n+\mathcal{E}_j)}+\frac{iN}{T}N_L(\bar{\rho}^2-q_0)\bar{\lambda}\\
    =&-N\sum_{\bold{k}\in \text{mBZ}}\sum_{j=1}^{2+4N_G} \ln^{(1+e^{\mathcal{E}_j/T})}+s\frac{iN}{T}N_L(\bar{\rho}^2-q_0)\bar{\lambda},
    \end{aligned}	
\end{equation}
\begin{equation}
	\begin{aligned}
A_0=
\begin{bmatrix}
 \begin{array}{ccccc}h^{(c)}(\bold{k}+\bold{G_1}) & 0 & \cdots & 0 &\bar{\rho} V(\bold{k}+\bold{G_1}) \\
0 & h^{(c)}(\bold{k}+\bold{G_2}) & \cdots & 0 & \bar{\rho} V(\bold{k}+\bold{G_2})\\ 
 \vdots & \vdots & \ddots & \vdots & \vdots\\
 0 & 0 & \cdots & h^{(c)}(\bold{k}+\bold{G_{N_{G}}})  & \bar{\rho} V(\bold{k}+\bold{G_{N_G}})\\
(\bar{\rho} V(\bold{k}+\bold{G_1}))^{\dagger} & (\bar{\rho} V(\bold{k}+\bold{G_2}))^{\dagger} & \cdots & (\bar{\rho} V(\bold{k}+\bold{G_{N_G}}))^{\dagger}&h^{(f)}(\bold{k})+i\bar{\lambda}\sigma_0\\
\end{array}
\end{bmatrix},
\end{aligned}
\end{equation}
we relabel
\begin{equation}
\begin{aligned}
\hat{h}^{(c)}= \left(\begin{array}{cccc}h^{(c)}(\bold{k}+\bold{G_1}) & 0 & \cdots & 0 \\
0 & h^{(c)}(\bold{k}+\bold{G_2}) & \cdots & 0 \\ 
 \vdots & \vdots & \ddots & \vdots \\
 0 & 0 & \cdots & h^{(c)}(\bold{k}+\bold{G_{N_{G}}}) \ \end{array}\right)
\end{aligned}
\end{equation}
and
\begin{equation}
\begin{aligned}
\hat{V}= \left(\begin{array}{cccc}\bar{\rho} V(\bold{k}+\bold{G_1}) & \bar{\rho} V(\bold{k}+\bold{G_2}) & \cdots & \bar{\rho} V(\bold{k}+\bold{G_{N_G}})  \ \end{array}\right).
\end{aligned}
\end{equation}
%

So, we can rewrite
\begin{equation}
	\begin{aligned}
A_0=
\begin{bmatrix}
 \begin{array}{ccccc}\hat{h}^{(c)} & \hat{V}  \\
\hat{V}^{\dagger} & h^{(f)}+i\bar{\lambda}\sigma_0 \\ 
\end{array}
\end{bmatrix},
\end{aligned}
\end{equation}

After applying $\partial S_{0}/\partial \bar{\rho}=0$ and $\partial S_{0}/\partial \bar{\lambda}=0$, 
 we have the saddle point mean field equations as follows
\begin{equation}
    \bar{\lambda}=-i\frac{T}{N_{L}}\sum_{\bold{k}\in \text{mBZ}}\sum_{\omega_n}\frac{1}{\det{(-i\omega_n +A_0)}}\frac{\partial(\det(-i\omega_n+A_0))}{(2\bar{\rho})\partial\bar{\rho}},
\end{equation}

\begin{equation}
      q_0-\bar{\rho}^{2}=i\frac{T}{N_{L}}\sum_{\bold{k}\in \text{mBZ}}\sum_{\omega_n}	\frac{1}{\det{(-i\omega_n +A_0)}}\frac{\partial(\det(-i\omega_n+A_0))}{\partial\bar{\lambda}}.
\end{equation}

Using the following formula
\begin{equation}
\frac{\partial{(\det{(C}))}}{\partial t}=(\det{(C}))\cdot\Tr((C)^{-1}\frac{\partial{C}}{\partial t}),
\end{equation}
where $C$ is a matrix and $t$ is a scalar.
We have
\begin{equation}
\frac{\partial{(\det{(-i\omega_n+A_0}))}}{\partial \bar{\rho}}=(\det{(-i\omega_n+A_0}))\cdot\Tr((-i\omega_n+A_0)^{-1}\frac{\partial{(-i\omega_n+A_0)}}{\partial \bar{\rho}})
\end{equation}
and
\begin{equation}
\frac{\partial{(\det{(-i\omega_n+A_0}))}}{\partial \bar{\lambda}}=(\det{(-i\omega_n+A_0}))\cdot\Tr((-i\omega_n+A_0)^{-1}\frac{\partial{(-i\omega_n+A_0)}}{\partial \bar{\lambda}})
\end{equation}.

Then the mean-field equations become
\begin{equation}
     \bar{\lambda}=-i\frac{T}{N_{L}}\sum_{\bold{k}\in \text{mBZ}}\sum_{\omega_n}\Tr((-i\omega_n+A_0)^{-1}\frac{\partial{(-i\omega_n+A_0)}}{(2\bar{\rho})\partial \bar{\rho}}),
\end{equation}

\begin{equation}
      q_0-\bar{\rho}^{2}=i\frac{T}{N_{L}}\sum_{\bold{k}\in \text{mBZ}}\sum_{\omega_n}	\Tr((-i\omega_n+A_0)^{-1}\frac{\partial{(-i\omega_n+A_0)}}{\partial \bar{\lambda}}).
\end{equation}
We define
\begin{equation}
	\begin{aligned}
A^{\bar{\rho}}_0=\frac{\partial{(-i\omega_n+A_0)}}{\partial \bar{\rho}}=
\begin{bmatrix}
 \begin{array}{ccccc}0 & 0 & \cdots & 0 & V(\bold{k}+\bold{G_1}) \\
0 & 0 & \cdots & 0 &  V(\bold{k}+\bold{G_2})\\ 
 \vdots & \vdots & \ddots & \vdots & \vdots\\
 0 & 0 & \cdots & 0  &  V(\bold{k}+\bold{G_{N_G}})\\
( V(\bold{k}+\bold{G_1}))^{\dagger} & ( V(\bold{k}+\bold{G_2}))^{\dagger} & \cdots & ( V(\bold{k}+\bold{G_{N_G}}))^{\dagger}&0\\
\end{array}
\end{bmatrix}
\end{aligned}
\end{equation}
and
\begin{equation}
	\begin{aligned}
A^{\bar{\lambda}}_0=\frac{\partial{(-i\omega_n+A_0)}}{\partial \bar{\lambda}}=
\begin{bmatrix}
 \begin{array}{ccccc}0 & 0 & \cdots & 0 &0 \\
0 & 0 & \cdots & 0 & 0\\ 
 \vdots & \vdots & \ddots & \vdots & \vdots\\
 0 & 0 & \cdots & 0 & 0\\
0 & 0 & \cdots & 0 & i\sigma_0\\
\end{array}
\end{bmatrix}.
\end{aligned}
\end{equation}
The mean field equations are written as
\begin{equation}
     \bar{\lambda}=-i\frac{T}{N_{L}}\sum_{\bold{k}\in \text{mBZ}}\sum_{\omega_n}\Tr((-i\omega_n+A_0)^{-1}\frac{A^{\bar{\rho}}_0}{2\bar{\rho}}),
\end{equation}

\begin{equation}
      q_0-\bar{\rho}^{2}=i\frac{T}{N_{L}}\sum_{\bold{k}\in \text{mBZ}}\sum_{\omega_n}	\Tr((-i\omega_n+A_0)^{-1}A^{\bar{\lambda}}_0).
\end{equation}

Actually, we can get the above equations from the action directly before we do $\Tr\ln=\ln\det$. We introduce an invertible matrix $P$ to diagonal the matrix $A_0$. Then $P$ can be constructed by the eigenvectors of $A_0$ and $P=(c_1,c_2,\cdots,c_{(2+4N_G)})_{2+4N_G\times2+4N_G}$, where $c_j$ are eigenvectors of $A_0$. Since
\begin{equation}
	\begin{aligned}
	&\Tr((-i\omega_n+A_0)^{-1}A^{\bar{\rho}}_0)=\Tr(P^{-1}(-i\omega_n+A_0)^{-1}PP^{-1}A^{\bar{\rho}}_0P)=\sum^{2+4N_G}_{j}((-i\omega_n+\mathcal{E}_j)^{-1}(P^{-1}A^{\bar{\rho}}_0P)_{jj})\\
    \end{aligned}
\end{equation}
and
\begin{equation}
	\begin{aligned}
	&\Tr((-i\omega_n+A_0)^{-1}A^{\bar{\lambda}}_0)=\Tr(P^{-1}(-i\omega_n+A_0)^{-1}PP^{-1}A^{\bar{\lambda}}_0P)=\sum^{2+4N_G}_{j}((-i\omega_n+\mathcal{E}_j)^{-1}(P^{-1}A^{\bar{\lambda}}_0P)_{jj})\\
    \end{aligned},
\end{equation}
then the mean-field equations become
\begin{equation}
    \bar{\lambda}\bar{\rho}=\frac{i}{2}\frac{T}{N_{L}}\sum_{\bold{k}\in \text{mBZ}}\sum_{\omega_n}\sum^{2+4N_G}_{j}\frac{(P^{\dagger}A^{\bar{\rho}}_0P)_{jj}}{i\omega_n-\mathcal{E}_j},
\end{equation}

\begin{equation}
      q_0-\bar{\rho}^{2}=-i\frac{T}{N_{L}}\sum_{\bold{k}\in \text{mBZ}}\sum_{\omega_n}	\sum^{2+4N_G}_{j}\frac{(P^{\dagger}A^{\bar{\lambda}}_0P)_{jj}}{i\omega_n-\mathcal{E}_j}.
\end{equation}
Note that $P$ is an unitary matrix and $P^{-1}=P^{\dagger}$. We also have the third mean field equation,
\begin{equation}
     n_{t}=q_0-\bar{\rho}^{2}+\frac{T}{N_{L}}\sum_{\bold{k}\in \text{mBZ}}\sum_{\omega_n}	\sum^{2+4N_G}_{j}\frac{(P^{\dagger}A^{c}_0P)_{jj}}{i\omega_n-\mathcal{E}_j}-2N_G,
\end{equation}
where $n_t=\text{constant}$ is the total number of flat band and conduction band electrons and we set $n_t=1.66$ below. We define

\begin{equation}
	\begin{aligned}
A^{c}_0=
\begin{bmatrix}
 \begin{array}{ccccc}\id_{4\times 4} & 0 & \cdots & 0 &0 \\
0 & \id_{4\times 4} & \cdots & 0 & 0\\ 
 \vdots & \vdots & \ddots & \vdots & \vdots\\
 0 & 0 & \cdots & \id_{4\times 4} & 0\\
0 & 0 & \cdots & 0 & 0_{2\times2}\\
\end{array}
\end{bmatrix}.
\end{aligned}
\end{equation}

After summing over the Matsubara frequency $\omega_n$, we have 

\begin{equation}
     \bar{\lambda}\bar{\rho}=\frac{i}{2N_{L}}\sum_{\bold{k}\in \text{mBZ}}\sum^{2+4N_G}_{j}(P^{\dagger}A^{\bar{\rho}}_0P)_{jj}\cdot n_{F}(\mathcal{E}_j),
\end{equation}

\begin{equation}
      q_0-\bar{\rho}^{2}=-\frac{i}{N_{L}}\sum_{\bold{k}\in \text{mBZ}}\sum^{2+4N_G}_{j}(P^{\dagger}A^{\bar{\lambda}}_0P)_{jj}\cdot n_F(\mathcal{E}_j),
\end{equation}

\begin{equation}
      n_t=q_0-\bar{\rho}^{2}+\frac{1}{N_{L}}\sum_{\bold{k}\in \text{mBZ}}\sum^{2+4N_G}_{j}(P^{\dagger}A^{c}_0P)_{jj}\cdot n_F(\mathcal{E}_j)-2N_G,
\end{equation}

where $n_F(\epsilon)=\frac{1}{\exp(\epsilon/T)+1}$ is the Fermi-Dirac distribution. Note that $\mathcal{E}_j(\bar{\lambda},\bar{\rho},\mu,\bold{k})$ is numerically calculated and depends on the chemical potential $\mu$ and momentum $\bold{k}$ with $\bold{k}\in \text{mBZ}$. 

\section{Green Functions}
Now, let us think about the Green functions.
\begin{equation}
G_0^{-1}=i\omega_n - A_0	.
\end{equation}
So, 
\begin{equation}
	\begin{aligned}
G^{ff}_0&=-i\frac{T}{N_{L}}\sum_{\bold{k}\in \text{mBZ}}\sum_{\omega_n}	\sum^{2+4N_G}_{j}\frac{(P^{\dagger}A^{\bar{\lambda}}_0P)_{jj}}{i\omega_n-\mathcal{E}_j},
\end{aligned}
\end{equation}
\begin{equation}
	\begin{aligned}
G^{cc}_0&=\frac{T}{N_{L}}\sum_{\bold{k}\in \text{mBZ}}\sum_{\omega_n}	\sum^{2+4N_G}_{j}\frac{(P^{\dagger}A^{c}_0P)_{jj}}{i\omega_n-\mathcal{E}_j},
\end{aligned}
\end{equation}
\begin{equation}
	\begin{aligned}
G^{cf}_0&=\frac{1}{2}\frac{T}{N_{L}}\sum_{\bold{k}\in \text{mBZ}}\sum_{\omega_n}\sum^{2+4N_G}_{j}\frac{(P^{\dagger}A^{\gamma}_0P)_{jj}}{i\omega_n-\mathcal{E}_j}
\end{aligned}
\end{equation}
\begin{equation}
	\begin{aligned}
G^{fc}_0&=\frac{1}{2}\frac{T}{N_{L}}\sum_{\bold{k}\in \text{mBZ}}\sum_{\omega_n}\sum^{2+4N_G}_{j}\frac{(P(A^{\gamma}_0)^{\dagger}P^{\dagger})_{jj}}{i\omega_n-\mathcal{E}_j},
\end{aligned}
\end{equation}
where 
\begin{equation}
	\begin{aligned}
A^{\gamma}_0=
\begin{bmatrix}
 \begin{array}{ccccc}0 & 0 & \cdots & 0 & \mathbb{1}^{0}_{4\times2} \\
0 & 0 & \cdots & 0 &  \mathbb{1}^{0}_{4\times2}\\ 
 \vdots & \vdots & \ddots & \vdots & \vdots\\
 0 & 0 & \cdots & 0  &  \mathbb{1}^{0}_{4\times2}\\
\mathbb{1}^{0}_{2\times4} & \mathbb{1}^{0}_{2\times4} & \cdots & \mathbb{1}^{0}_{2\times4}&0\\
\end{array}
\end{bmatrix}
\end{aligned}
\end{equation}
and
\begin{equation}
	\begin{aligned}
\mathbb{1}^{0}_{4\times2}=
\begin{bmatrix}
 \begin{array}{ccccc}1 & 0  \\
0 & 1 \\ 
 0 & 0 \\
0 & 0 \\
\end{array}
\end{bmatrix}
\end{aligned}
\end{equation}
with
\begin{equation}
	\begin{aligned}
\mathbb{1}^{0}_{2\times4}=
\begin{bmatrix}
 \begin{array}{ccccc}1 & 0 & 0 & 0 \\
0 & 1 & 0 & 0\\ 
\end{array}
\end{bmatrix}.
\end{aligned}
\end{equation}

\section{Numerics}
In this section, we give the data details of the numerics. We also give more numerical details. First of all, the three mean field equations Eq.~\ref{eqS1},~\ref{eqS2}, and ~\ref{eqS1} are coupled with each other. As one approaches the Kondo region which means $\rho\approx0$, the Eq.~\ref{eqS1} will be automatically satisfied. One no longer needs to consider Eq.~\ref{eqS1} in the Kondo region. Second, to solve the three equations self-consistently, we numerically calculate $\mathcal{E}_j(\bar{\lambda},\bar{\rho},\mu_f,\bold{k})$ by diagonalizing $A_0$ for each value of $\bar{\lambda}$, $\bar{\rho}$, and $\mu_f$. We set a certain value of temperature $T$ and set $\mu_c=\mu_f$ then go through the parameters regions $\rho \in[0,0.5]$, $\mu_f \in [-100,100]$ meV, and $i\lambda \in [-100,100]$ meV. Since we set the interaction $U=\infty$, $Q=1$, so we have $q_0=1/4$. We find the solutions to make the $errs\approx0$. There exist two solutions: one is positive $\mu_f$, and another is negative $\mu_f$. We note that momentum $\bold{k}$ belongs to the first $\text{mBZ}$ with $\bold{k}\in \text{mBZ}$. To check the convergence, we perform $N_G$ from $3$ to $37$ and it turns out $3~\text{mBZs}$ are good enough for the convergence. 
\begin{equation}
    \bar{\rho}\bar{\lambda}=\frac{i}{2 N_{L}}\sum_{\bold{k}\in \text{mBZ}}\sum^{2+4N_G}_{j}(P^{\dagger}A^{\bar{\rho}}_0P)_{jj}\cdot n_{F}(\mathcal{E}_j),\label{eqS1}
\end{equation}
\begin{equation}
      q_0-\bar{\rho}^{2}=-\frac{i}{N_{L}}\sum_{\bold{k}\in \text{mBZ}}\sum^{2+4N_G}_{j}(P^{\dagger}A^{\bar{\lambda}}_0P)_{jj}\cdot n_F(\mathcal{E}_j),\label{eqS2}
\end{equation}
\begin{equation}
      n_t=q_0-\bar{\rho}^{2}+\frac{1}{N_{L}}\sum_{\bold{k}\in \text{mBZ}}\sum^{2+4N_G}_{j}(P^{\dagger}A^{c}_0P)_{jj}\cdot n_F(\mathcal{E}_j)-2N_G,\label{eqS3}
\end{equation}

The following data are for the self-consistent solutions of $n_t=0.8q_0$. As $\rho	\approx0.24$ at $T=24K$, $n_c\approx0$ which means $T=24K$ is the Kondo temperature for $n_t=0.8q_0$. The reason why $\rho$ is not zero is because we set $n_t=0.8q_0$, this means parts of $n_c$ are in higher energy conduction bands.
\begin{table}[h!]
  \begin{center}
    \caption{Temperature Dependence for $n_t=0.8q_0$.}
    \label{tab:table1}
    \begin{tabular}{l|c|c|c|c|c|c|c|c|c|r} 
      \hline
      \hline
      $n_t$ & 1/4$\times$0.8 & 1/4$\times$0.8 & 1/4$\times$0.8  & 1/4$\times$0.8 & 1/4$\times$0.8 & 1/4$\times$0.8 & 1/4$\times$0.8 & 1/4$\times$0.8 & 1/4$\times$0.8 & 1/4$\times$0.8\\
      \hline
      $q_0$ & 1/4 & 1/4 & 1/4& 1/4 & 1/4 & 1/4 & 1/4 & 1/4 & 1/4 & 1/4\\
      \hline   
      $T$ & $0.01K$ & $0.4K$ & $0.8K$ & $1.2K$ & $1.6K$ & $2.0K$ & $4.0K$ & $6.0K$ & $8.0K$ & $10.0K$\\
      \hline
      $\rho$ & 0.4 & 0.4 & 0.41 & 0.4 & 0.4 & 0.4 & 0.4 & 0.39 & 0.4 & 0.39\\
      \hline
      $\mu_f$ & 19 & 19 & 20 & 19 & 20 & 19 & 19 & 18 & 20 & 18\\
      \hline
      $i\lambda$ & $22$ & $22$ & $23$ & $22$ & $23$ & $22$ & $22$ & $21$ & $24$ & $22$   \\
      \hline
      $n_f$ & $0.09$ & $0.09$ & $0.11$ & $0.09$ & $0.09$ & $0.09$ & $0.09$ & $0.09$ & $0.09$ & $0.09$\\
      \hline
      $n_c$ & $0.11$ & $0.11$ & $0.09$ & $0.11$ & $0.11$ & $0.11$ & $0.11$ & $0.11$ & $0.11$ & $0.11$\\
      \hline
      $errs$ & $0.015$ & $0.015$ & $0.0084$ & $0.011$ & $0.01$ & 0.0087  & 0.013 & 0.0081 & 0.0058 & 0.0052\\
      \hline              
    \end{tabular}
  \end{center}
\end{table}

\begin{table}[h!]
  \begin{center}
    \caption{Temperature Dependence for $n_t=0.8q_0$.}
    \label{tab:table1}
    \begin{tabular}{l|c|c|c|c|c|c|c|c|c|c|r} 
      \hline
      \hline
      $n_t$ & 1/4$\times$0.8 & 1/4$\times$0.8 & 1/4$\times$0.8  & 1/4$\times$0.8 & 1/4$\times$0.8 & 1/4$\times$0.8 & 1/4$\times$0.8  & 1/4$\times$0.8 & 1/4$\times$0.8 & 1/4$\times$0.8 & 1/4$\times$0.8\\
      \hline
      $q_0$ & 1/4 & 1/4 & 1/4& 1/4 & 1/4 & 1/4 & 1/4& 1/4 & 1/4 & 1/4 & 1/4\\
      \hline   
      $T$  & $12K$ & $14K$ & $16K$ & $18K$ & $20K$ & $22K$ & $24K$ & $26K$ & $28K$ & $29K$ & $30K$\\
      \hline
      $\rho$ & 0.4 & 0.39 & 0.37 & 0.35 & 0.32 & 0.28 & 0.24 & 0.19 & 0.13  & 0.09 & 0\\
      \hline
      $\mu_f$ & 20 & 18 & 15 & 12 & 6 & 3 & 0 & -1 & -3 & -3.4  & -4\\
      \hline
      $i\lambda$ & $25$ & $23$ & $20$ & $17$ & $11$ & 8 & $5$ & $4$ & 2  & 1.6  & 1\\
      \hline
      $n_f$ & $0.09$ & $0.09$ & $0.11$ & $0.13$ & $0.15$ & $0.17$ & $0.19$ & $0.21$ & $0.23$ & $0.24$ & $0.25$\\
      \hline
      $n_c$ & $0.11$ & $0.11$ & $0.09$ & $0.07$ & $0.05$ & $0.03$ & $0.01$ & $-0.01$ & $-0.03$ & $-0.04$ & $-0.05$ \\
      \hline
      $errs$ & $0.0047$ & $0.0071$ & $0.0031$ & $0.0032$ & $0.0043$ & 0.0069 & $0.0059$ & $0.0064$ & 0.0047 & 0.0024 & 0.0034\\
      \hline              
    \end{tabular}
  \end{center}
\end{table}


\end{document}